\documentstyle[11pt,aaspp4]{article}
\slugcomment{ Accepted by Astron. J.}

\lefthead{Benedict et al.}
\righthead{Photometry of Proxima Centauri and Barnard's Star}

\begin{document}

\title{Photometry of Proxima Centauri and Barnard's Star\\ Using HST Fine
Guidance Sensor 3: \\
A Search for Periodic Variations\footnote{Based on observations made with
the NASA/ESA Hubble Space Telescope, obtained at the Space Telescope
Science Institute, which is operated by the
Association of Universities for Research in Astronomy, Inc., under NASA
contract NAS5-26555} }

\author{ G.\ Fritz Benedict\altaffilmark{2}, Barbara McArthur\altaffilmark{2},
E.\ Nelan\altaffilmark{3}, D. Story\altaffilmark{9}, A.\ L.\
Whipple\altaffilmark{10},
P.~J.~Shelus\altaffilmark{2}, W.\ H.\ Jefferys\altaffilmark{4},
 P.D. Hemenway\altaffilmark{5},	Otto G.\ Franz\altaffilmark{6}, L.\ H.\
Wasserman\altaffilmark{6},	R. L.
Duncombe\altaffilmark{11},Wm.~van~Altena\altaffilmark{7}, and L.W.\
Fredrick\altaffilmark{8}}
\altaffiltext{2}{McDonald Observatory, University of Texas}
 \altaffiltext{3}{Space Telescope Science Institute}
\altaffiltext{4}{Astronomy Dept., University of Texas}
 \altaffiltext{5}{University of Rhode Island}
 \altaffiltext{6}{Lowell Observatory}
 \altaffiltext{7}{Astronomy Dept., Yale University}
\altaffiltext{8}{Astronomy Dept., University of Virginia}
\altaffiltext{9}{McDonald Observatory, University of Texas. Now at Jackson
\& Tull.}
\altaffiltext{10}{McDonald Observatory, University of Texas. Now at
Allied-Signal Aerospace.}
\altaffiltext{11}{Aerospace Engineering, University of Texas}

% Notice that each of these authors has alternate affiliations, which
% are identified by the \altaffilmark after each name.  The actual alternate
% affiliation information is typeset in footnotes at the bottom of the
% first page, and the text itself is specified in \altaffiltext commands.
% There is a separate \altaffiltext for each alternate affiliation
% indicated above.

% The abstract environment prints out the receipt and acceptance dates
% if they are relevant for the journal style.  For the aasms style, they
% will print out as horizontal rules for the editorial staff to type
% on, so long as the author does not include \received and \accepted
% commands.  This should not be done, since \received and \accepted dates
% are not known to the author.

\begin{abstract}
We have observed Proxima Centauri and Barnard's Star with {\it Hubble Space Telescope} Fine Guidance Sensor 3.
Proxima Centauri  exhibits small-amplitude, periodic
photometric variations.  Once several sources of systematic photometric
error are corrected, we obtain 2 milli-magnitude internal photometric
precision.
We identify
two distinct behavior modes over the past four years: higher amplitude,
longer period; smaller
amplitude, shorter period. Within the errors one period ($P \sim 83^{d}$)
 is twice the other. Barnard's Star shows very weak evidence for periodicity on a
timescale of approximately 130 days. If we interpret these periodic
phenomena as rotational modulation of star spots, we identify three
discrete spots on
Proxima Cen and possibly one spot on Barnard's Star. We find that the disturbances
change
significantly on time scales as short as one rotation period.
\end{abstract}

% The different journals have different requirements for keywords.  The
% keywords.apj file, found on aas.org in the pubs/aastex-misc directory,
% contains a list of keywords used with the ApJ and Letters.  These are
% usually assigned by the editor, but authors may include them in their
% manuscripts if they wish.


% That's it for the front matter.  On to the main body of the paper.
% We'll only put in tutorial remarks at the beginning of each section
% so you can see entire sections together.

% In the first two sections, you should notice the use of the LaTeX \cite
% command to identify citations.  The citations are tied to the
% reference list via symbolic KEYs.  We have chosen the first three
% characters of the first author's name plus the last two numeral of the
% year of publication.  The corresponding reference has a \bibitem
% command in the reference list below.
%
% Please see the AASTeX manual for a more complete discussion on how to make
% \cite-\bibitem work for you.

\section{Introduction}

We present photometry of Proxima Cen and Barnard's Star, results ancillary
to our astrometric searches for
 planetary-mass companions (\cite{Ben97a}).  Our observations  were
obtained with Fine Guidance Sensor 3 (FGS 3), a two-axis, white-light
interferometer  aboard the {\it Hubble Space Telescope (HST)}. \cite{Bra91}
provide an overview of the
FGS 3 instrument and \cite{Ben94a} describe the astrometric capabilities of
FGS 3 and typical data
acquisition strategies.   \cite{Ben93} assessed  FGS 3 photometric
qualities and presented the first
evidence for periodic variability of Proxima Cen. This latter result was
based on  212 days of
monitoring.  Subsequent data exhibited a period of variation very nearly
twice the original
(\cite{Ben94b}).
Since that report, we have obtained  14 additional data sets for Proxima
Cen and 12 new sets for
Barnard's Star. The primary value of these observations lies in their
precision, not
in their temporal span or aggregate numbers. We have previously determined
that a 90 sec observation obtained with FGS 3 has a $1-\sigma$ precision of
0.001
magnitude at $V = 11$ (\cite{Ben93}), in the absence of systematic errors.

 In this paper we discuss the  data sets and assess systematic errors,
including background contamination and FGS position-dependent photometric
response. We also
present a revised photometric flat field. We then exhibit and analyze light
curves for Proxima Cen and
Barnard's Star. We find weak evidence for periodic variations in
the brightness of Barnard's Star. However, Proxima Cen exhibits significant periodic photometric variations,
with  changes in amplitude and/or period. We next interpret these
variations
as rotational modulation of chromospheric structure (star spots and/or plages),
and conclude with a brief comparison to other determinations of the
rotation rate of Proxima Cen. Tables~\ref{tbl-1} and
\ref{tbl-2} provide aliases and physical parameters for our two science targets.

We use the term `pickle' to describe the total field of view of the FGS.
The instantaneous field of view of FGS 3 is a
$5 \times 5$ arcsec square aperture.  Figure~\ref{fig-1} shows a finding
chart for the Barnard's Star reference frame in the  FGS 3 pickle as
observed on 6 August 1994. \cite{Ben93} contains a finding chart for the
Proxima Cen reference frame.

\section{Data Reduction} \label{dred}
\subsection{ The Data}
All position and brightness measurements from FGS 3 are comprised of series
of 0.025 sec samples (e.g.,
40Hz data rate), of between 20 and 120 sec or
$\sim600$ sec duration. Each FGS contains four
photomultipliers, two for each axis. We sum the output of all four to
produce our measurement, S, the average count per 0.025 sec sample,
obtained during the entire exposure.
 The coverage for both targets suffers from extended gaps, due to {\it HST}
pointing constraints
(described in \cite{Ben93}) and other scheduling difficulties. The filter
(F583W) has a bandpass
centered on 583 nm, with 234 nm FWHM.

For Proxima Cen the data now include 152 shorter exposures
secured over 4 years (March 1992 to October 1997) and 15 longer exposures
(July 1995 to July 1996).  Each
orbit contains from two to four exposures. The longest exposure times
pertain only to Proxima Cen
observations obtained within Continuous Viewing Zone (CVZ) orbits. These
specially scheduled orbits
permit $\sim90$ minutes on field, during which  Proxima Cen was not
occulted by the Earth.
Appendix 1.1 gives times of observation, exposure times, and average
counts, S,  for all Proxima Cen photometry.

Barnard's Star was monitored for three years (February 1993 to April 1996),
and observed three times during
each of 35 orbits. Exposures range between 24 and 123 seconds duration.
Appendix 1.2  gives times of observation, exposure times, and average
counts, S, for all Barnard's Star photometry.
\subsection{Background Light}
 We first noted that background contamination might be an issue while
assessing the use
 of astrometric reference stars for photometric flat-fielding. These stars
are typically far fainter
than the primary science targets. Using them to flat field the Proxima Cen
and Barnard's Star photometry introduced a strong one-year periodicity
(and considerable noise, since they are fainter stars). This problem was
not identified in
\cite{Ben93}, since we had access to data spanning less than two-thirds of
a year.   Figure~\ref{fig-2}
shows S for two faint reference stars in the Barnard
field plotted against angular distance from the Sun. These stars appear
brightest when closest to the Sun. Zodiacal light is a source whose
brightness
depends on the sun-target separation. The fitting function in
Figure~\ref{fig-2} is
\begin{equation}
I = A + B \sin(\frac{\theta}{2}), \label{ZODeqn}
\end{equation}
chosen to produce a minimum contribution at $\theta = 180 \arcdeg$. We find
$A = 137.1\pm 0.4 $ and $B= -4.1\pm 0.5$ counts per 25ms for an average
exposure time of 100 sec.  At a 60\arcdeg ~elongation the contamination
amounts to $V
= 22.5 \pm 0.3$ arcsec$^{-2}$. The Barnard field is at ecliptic latitude
$\beta = +27 \arcdeg$. From a tabulation in Allen
(1972)  we calculate a signal equivalent to V = 22.1 arcsec$^{-2}$ for
zodiacal light at $60\arcdeg$
 elongation and ecliptic latitude $\beta = +30 \arcdeg$. The agreement
supports our identification of this background source.

We present in Figure~\ref{fig-3}  an average S for these two Barnard
reference stars plotted as a function of time, uncorrected and corrected
for background. These data have been flat fielded using the time-dependent response function discussed in section~\ref{flat} (equation~\ref{FFeqn}).
Note the reduction in the amplitude of the scatter for the corrected
photometry. Presuming
Zodiacal Light as the source,  contamination levels are even less for the
Proxima Cen observations at ecliptic latitude $\beta = -44\arcdeg$,
introducing a maximum systematic
error of 0.0007 magnitude for a 100 sec observation. We conclude that
the effects of this component of the background are insignificant for
Proxima Cen and Barnard's Star photometry.

Should background determination become more important in the future, we
note that during an intra-orbit observation sequence the PMT are never
turned off.
 Hence, the
{\it HST} data archive contains PMT measurements taken during slews from
one star to the next. The astrometric reduction pipeline at the Space
Telescope Science Institute has been modified to provide these background
data automatically.

\subsection{Photometric Flat Fielding}
We explore two kinds of flat fielding; position- and time-dependent. We
first assess  whether or not
 flat-field corrections are necessary, and, if so, determine their
functional form.

\subsubsection{Position-dependent Photometric Response} \label{PDPR}
  Having discovered that background variations
contaminate the photometry of faint astrometric reference stars, we required an
alternative source for flat field data. To maintain the astrometric
calibration of FGS 3, a star field
in M35 has been measured roughly once per month for the last four years.
\cite{Whi95}  describe this continuing astrometric Long-term Stability
(LTSTAB) test. The field, on the ecliptic,
and, hence, always observed in one of two orientations (Fall or Spring)
flipped by $180 \arcdeg$,
contains bright stars for which background contamination is negligible.
However, an initial application
of a time-dependent flat field based on bright M35 stars also introduced a
strong one-year periodicity.

The positions of the three M35 LTSTAB stars within FGS 3 are shown in
Figure~\ref{fig-4} .   The
`eye' is bordered by the pickle edge at the two nominal rolls for this
field. The central
circle (diameter $\sim 3 \farcm 8$) is accessible by the FGS 3
instantaneous aperture for any HST roll.

\begin{figure}
\plotone{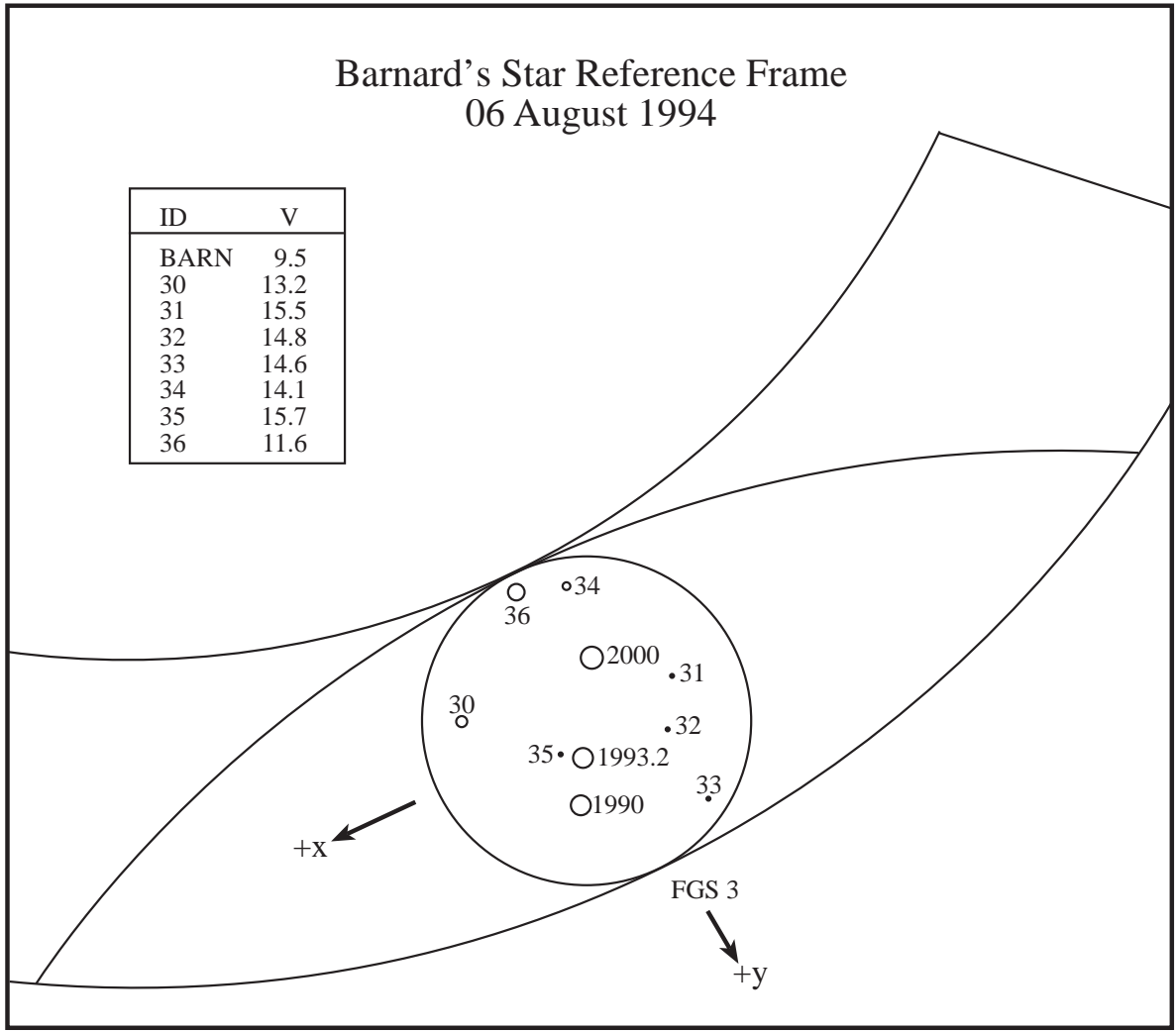}
\caption{The Barnard's Star field on 4 August 1994. North is at top, east
to the left. The pickle coordinate system (x, y) is indicated. The symbol
size is proportional
to the relative brightness of each reference star (listed). The central
circle (diameter $\sim 3 \farcm 8$) is accessible by the FGS 3
instantaneous aperture for any HST roll. Barnard's Star is labeled at three
epochs.} \label{fig-1}
\end{figure}

\begin{figure}
\plotone{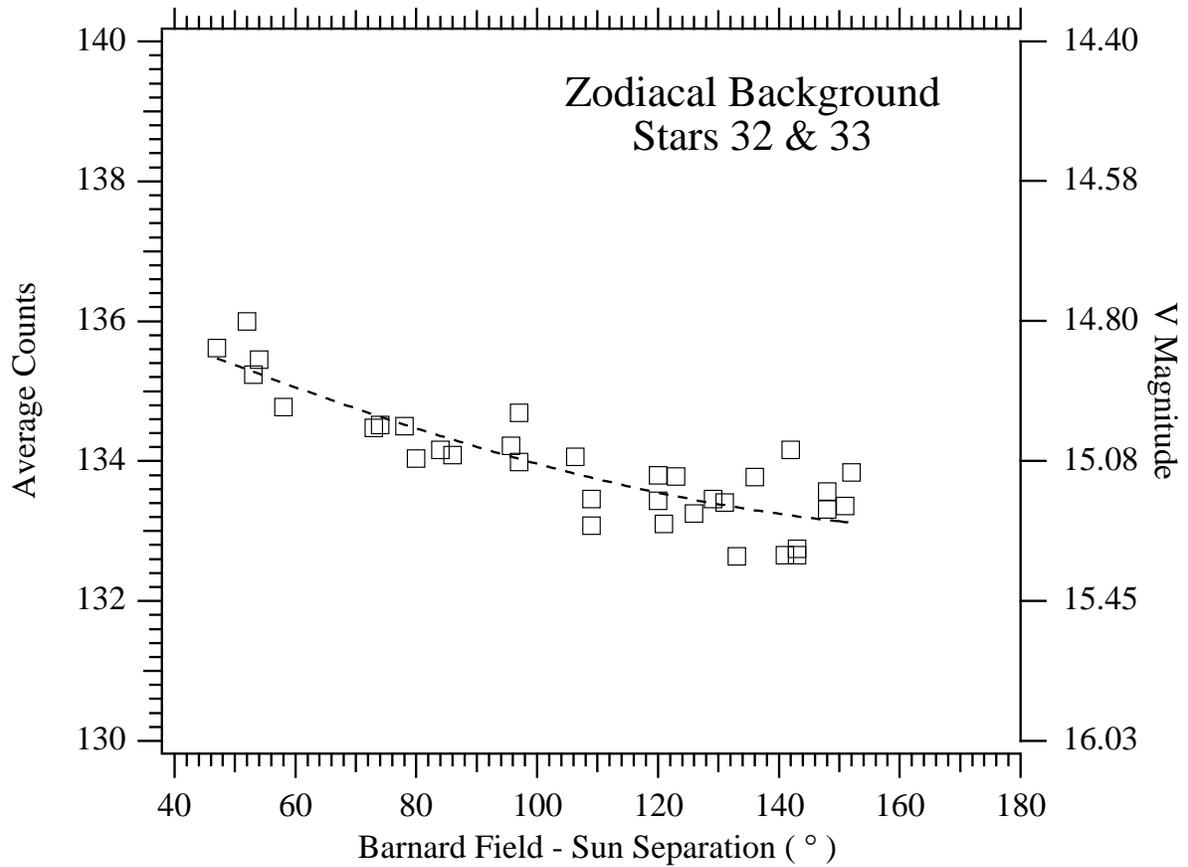}
\caption{S (average counts per 0.025 sec) for two faint astrometric
reference stars in
the Barnard's Star field (stars 32 and 33, Figure~\ref{fig-1}) vs. target -
Sun separation
in degrees. Fitting function is equation~\ref{ZODeqn}. } \label{fig-2}
\end{figure}

\begin{figure}
\plotone{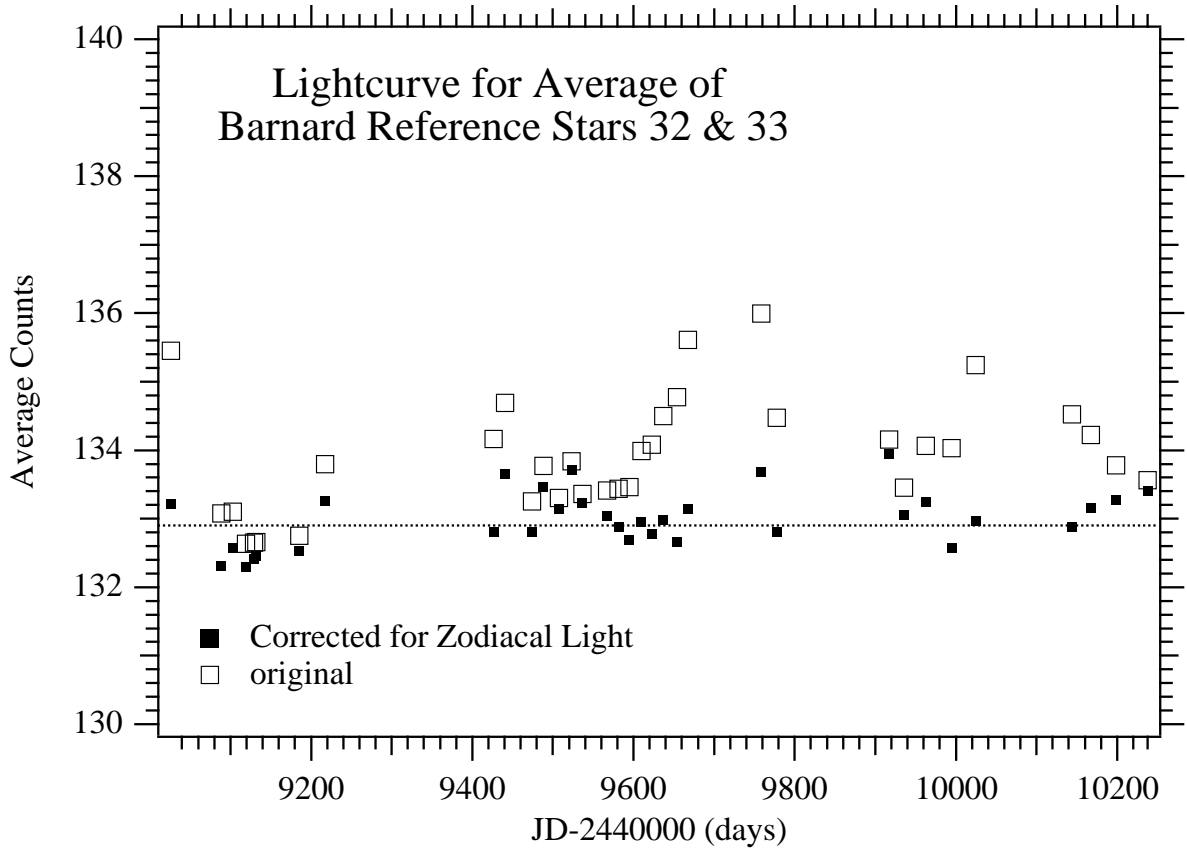}
\caption{Correcting for Zodiacal Light. Average intensity (S, counts per 0.025 sec) vs. modified
Julian Date for the average of two faint
astrometric reference stars in
the Barnard's Star field  (Figure~\ref{fig-2}). The horizontal line denotes
the mean brightness of the corrected data. Scatter is far less in the corrected data. Uncorrected and corrected values have been flat-fielded with equation
~\ref{FFeqn}} \label{fig-3}
\end{figure}

\begin{figure}
\plotone{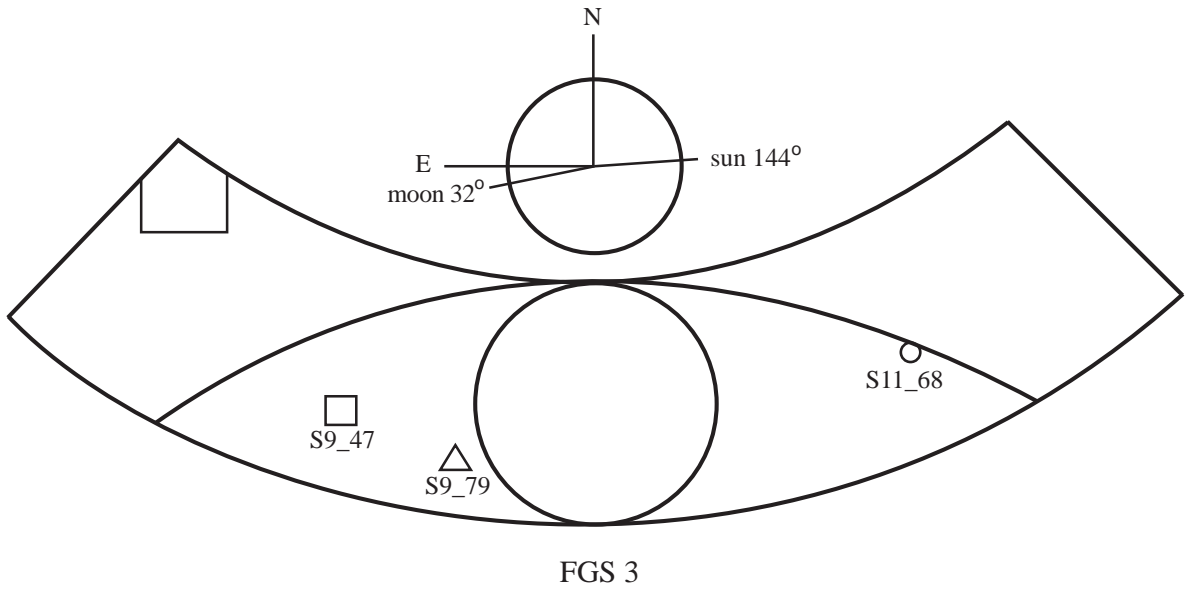}
\caption{Field of view of FGS 3 with position of three M35 stars. The V
magnitude
is given as part of the identification (e.g., S9\_ 47 has V = 9.47). }
\label{fig-4}
\end{figure}

\begin{figure}
\plotone{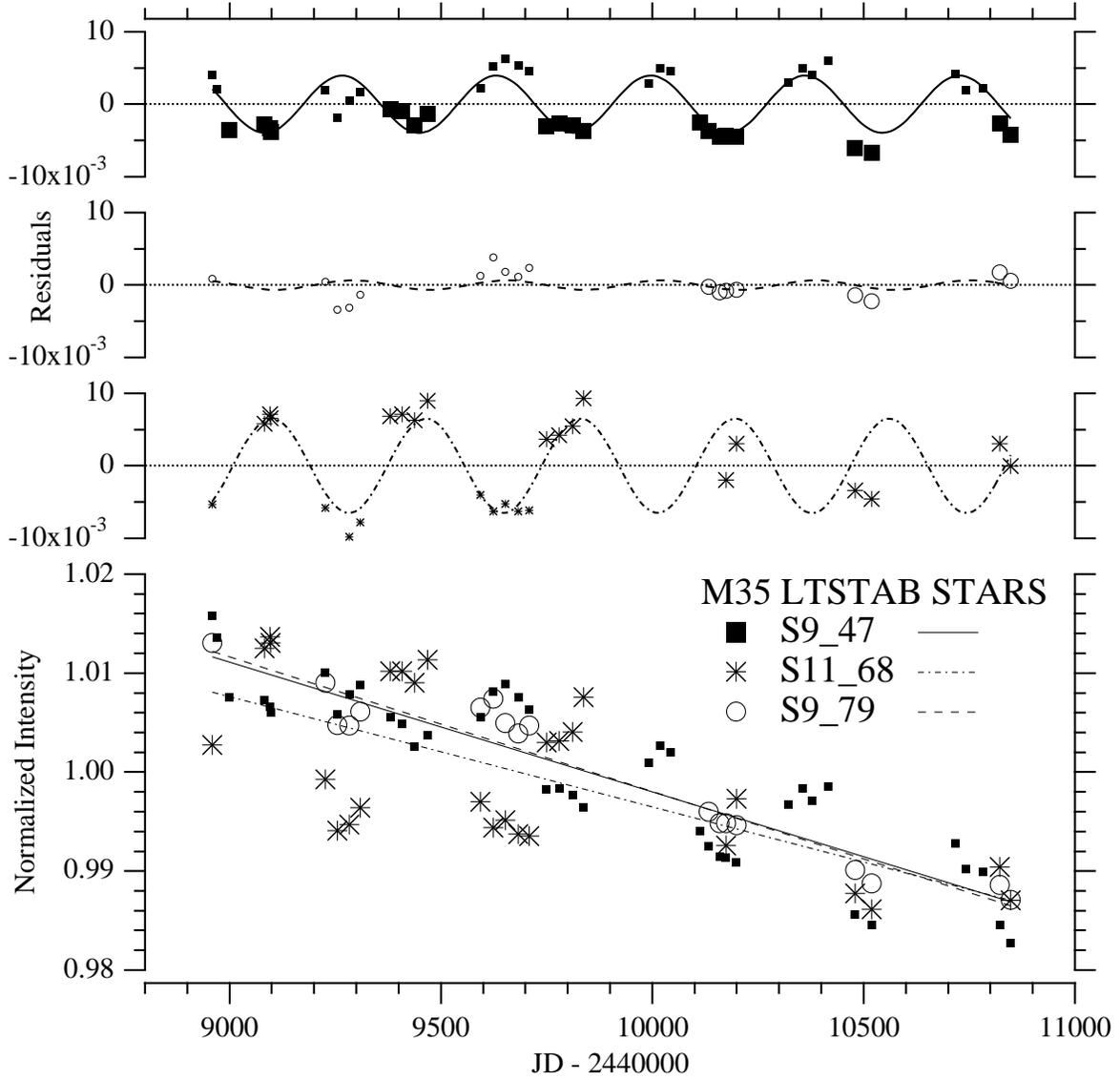}
\caption{Time-dependent photometric variations for three stars in M35
modeled as linear trends whose parameters (I and I') are given in Table~\ref{tbl-3}. 
The residuals in the top panels are size-coded to show the two LTSTAB
orientations (Fall = small, Spring = large) and
 fit with sine functions. Note differences in the variation
amplitude (parameter A in Table~\ref{tbl-3}).  } \label{fig-5}
\end{figure}

Figure~\ref{fig-5} (bottom) presents normalized intensities (I = S(t)/S$_{av}$,
where S$_{av}$ is determined from the entire run of data) for the three
LTSTAB stars as a function
of time.  The variation of each star has first been modeled by a linear trend. The parameters, intercept ($I_{o}$) and slope ($I'$), are given in Table~\ref{tbl-3}. The
resulting residuals (Figure~\ref{fig-5}, top) have been modeled with a sine wave.
while constraining $P = 365 \fd 25$ days. The residuals have a square-wave
periodic structure because, rather than a range of spacecraft rolls, there
are only two orientations.  The resulting parameters and errors are given
in
Table~\ref{tbl-3}. In Figure~\ref{fig-7} we plot the amplitude of this
side-to-side variation against
radial distance from the pickle center. For the M35 stars. the further the star from the pickle center, the larger the roll-induced variation. Figure~\ref{fig-7} includes several other
one year period amplitudes; a preliminary result for GJ748 ($V \sim 11.1$,
ecliptic lat $\beta \sim +23 \arcdeg$) always observed in the center of the
pickle, the Barnard reference star photometry from
Figure~\ref{fig-3} corrected for background, and photometry of the
brightest reference star in the Barnard field (Figure~\ref{fig-1}, star 36,
$V \sim 11.5$).  Figure~\ref{fig-7} suggests that within the inscribed
circle of
Figure~\ref{fig-1} ($r < 180\arcsec$),
 position-dependent photometric response variations  should be less than
0.002 magnitude.

We have also identified one high spatial frequency position-dependent flat
field component for FGS 3. Light curves for two of the Barnard reference
stars evidenced sudden decreases in brightness with subsequent return to
previous levels. The decrease for reference star 34  was 29\%; for star 36,
17\% . Shown
in Figure~\ref{fig-8}, both decreases occurred in the same location within FGS 3, very near the -Y edge. The  pickle X, Y of the
center of this area is (X, Y) = -25, 627. We estimate the size of the
low-sensitivity region to be $\sim
10 \times 10$ arcsec. Additionally,  Proxima Cen reference star
observations acquired one year prior to the Barnard's Star reference star
observations and within a few arcsec of this position showed no decrease,
providing additional evidence that FGS 3 is not suitable for wide-field,
precise faint star photometry.

Evidence that the photometric response may vary locally and randomly with time
dissuades us from mapping a position-dependent flat field over the entire
pickle.
However, for bright stars ($V < 11$) observed within $\sim 20$ arcsec of
the pickle center (Figure~\ref{fig-1}), these identified systematics should
produce very little effect. {\it All Proxima Cen and Barnard's Star
observations were secured
within 15 arcsec of the pickle center.}

\subsubsection{Time-dependent Photometric Response} \label{flat}

\begin{figure}
\plotone{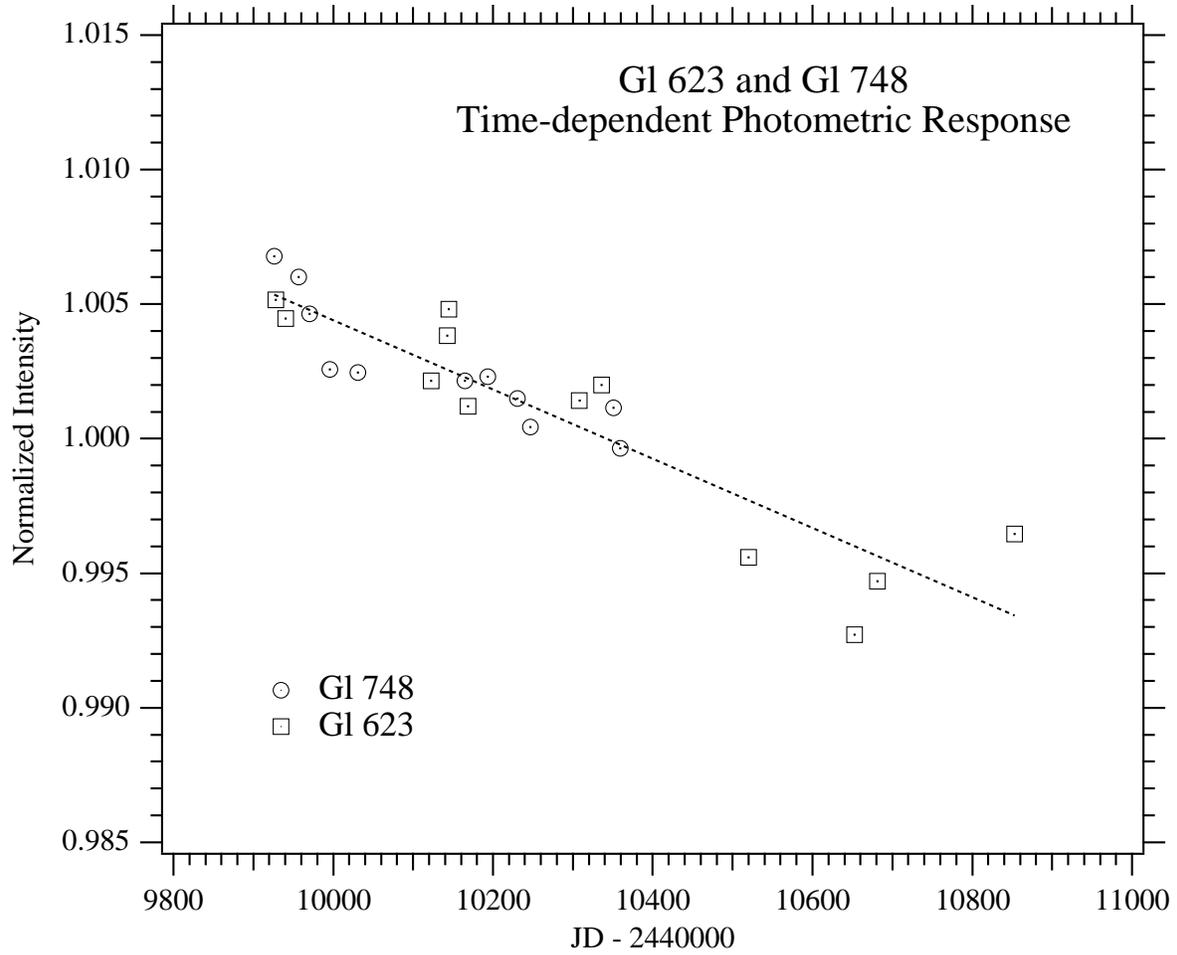}
\caption{Trend of normalized intensity for Gl 623  and
GJ 748. The fitted line parameters are given in Table~\ref{tbl-3}} \label{fig-6}
\end{figure}

\begin{figure}
\plotone{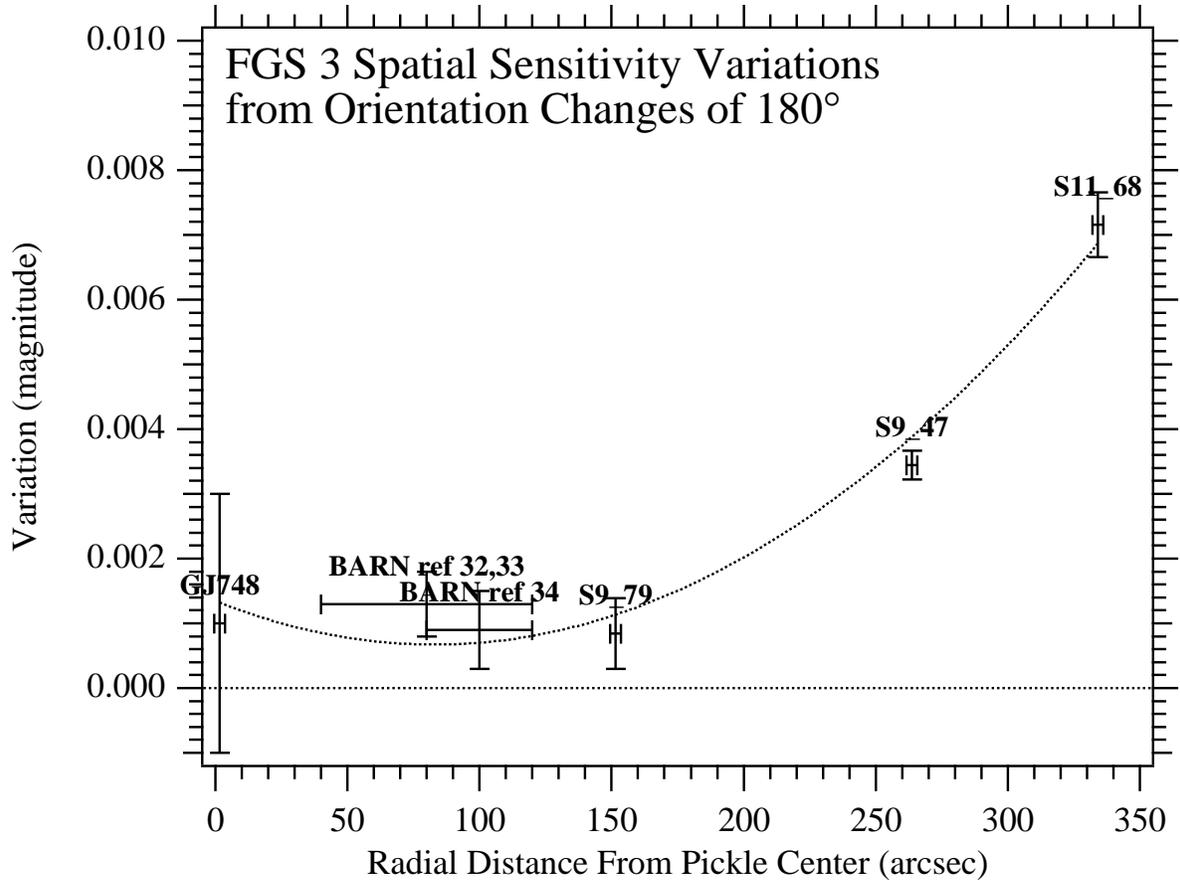}
\caption{Amplitude of side-to-side variation vs. distance from pickle center.
Targets are stars from M35 (location from Figure~\ref{fig-4}~and amplitude
from Figure~\ref{fig-5}), stars from the Barnard's Star reference frame
(Figure~\ref{fig-1}), and
GJ 748. Error bars along the X-axis indicate the radial range within the
pickle for all observations of the particular target.} \label{fig-7}
\end{figure}

\begin{figure}
\plotone{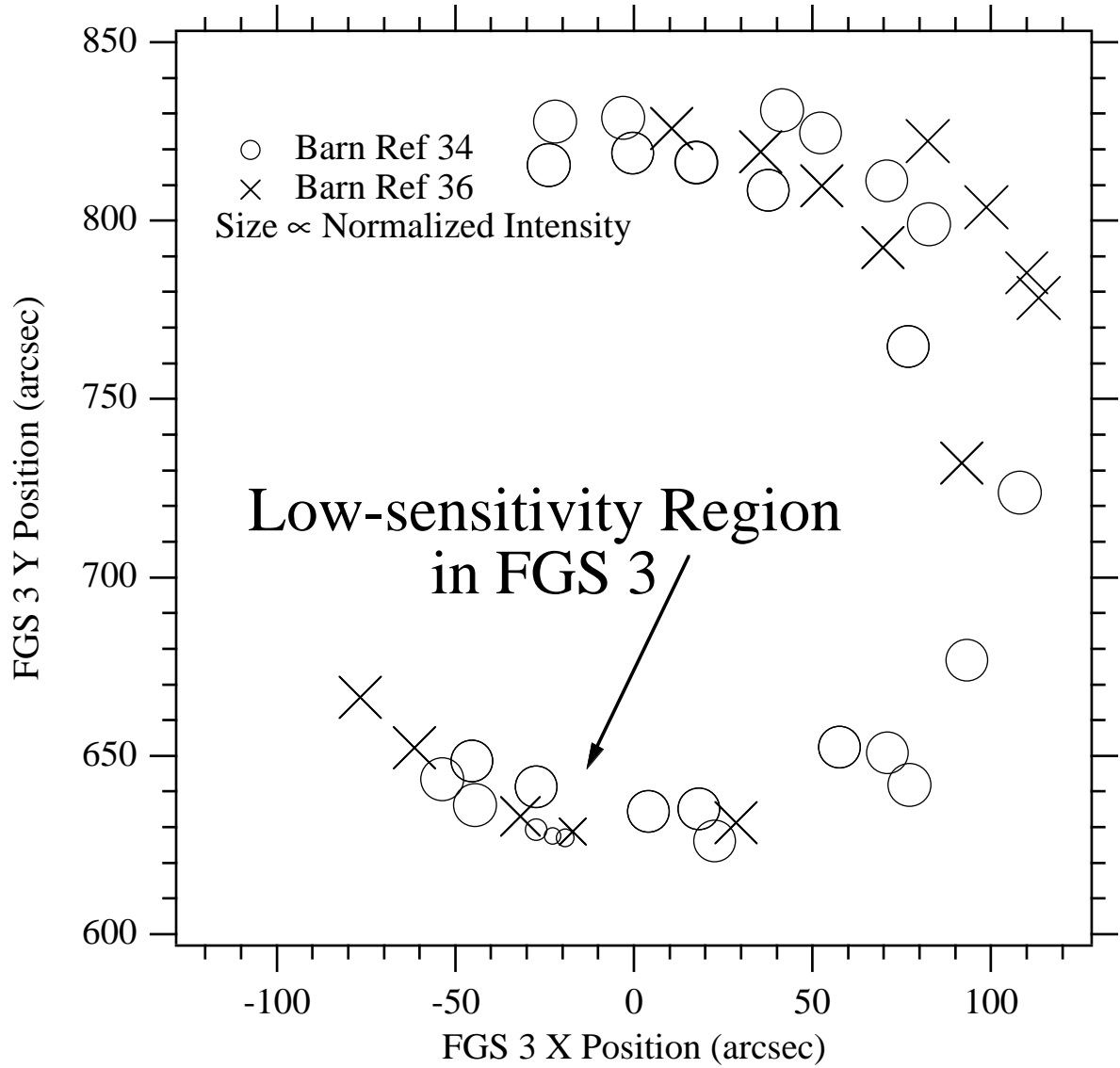}
\caption{Intensities of Barnard field reference stars 34 and 36 plotted
against pickle coordinates for all observation dates. The symbol size is
proportional to the normalized intensity. The low-sensitivity region is
clearly seen in the vicinity of (x, y) = (-25, 627).} \label{fig-8}
\end{figure}

Figure~\ref{fig-5} 
indicates that FGS 3 has become
less sensitive with age. For all three LTSTAB stars the linear trends
($I'$, Table~\ref{tbl-3}) agree within the errors. 
The apparent 1\% drop in sensitivity over 1000 days requires confirmation. 
 Figure~\ref{fig-6} presents the time varying normalized intensity
for two other astrometric program stars observed with FGS 3, GJ 623 and GJ 748. Both M dwarfs were observed in pickle center. Comparing the $I'$ in Table~\ref{tbl-3} and Table~\ref{tbl-4}, the rate of decline in brightness for GJ 623 and GJ 748 is identical (within the errors) to that seen in the M35 stars.

A final concern is that the rate of decline of PMT sensitivity might vary
with wavelength. The M35 stars (stars 547, 500, and 312 in the catalog of Cudworth, 1971) have $0.12 < B-V < 0.49$, while GJ 623 and GJ 748
have $B-V \simeq +1.5$. There appears to be no dependence on color.
 
The weighted average for five stars from three different fields yields
\begin{equation}
 FF = 1.131 \pm 0.006 + (1.30 \pm 0.06) \times 10^{-5} mJD  \label{FFeqn}
\end{equation}
  as the temporal photometric flat field for the pickle center.

As an additional test of the reality of this sensitivity decrease, we
note that the intensity data for the two astrometric reference stars in the Barnard's Star field shown in Figure~\ref{fig-3} have been flat-fielded with
equation~\ref{FFeqn}. Thus, a total of seven stars from four different fields show similar brightness trends, adequate evidence for a sensitivity loss 
in FGS 3.

\subsection{Photometric Calibration}
All magnitudes presented in this paper are provisional, since a final
calibration from F585W to $V$ is not
yet available.  If magnitudes are given, they are derived through
\begin{equation}
V = -2.5 \log( S ) + 20.0349
\end{equation}
with no color term, where S is the average count per 0.025 sec sample,
summed over all
four PMT. No results are based on these provisional calibrated magnitudes.
They are provided only as a convenience.

\subsection{Summary: Photometric Error and Photometric Precision}
We have identified sky background (Zodiacal Light), within-pickle response variations, and
time-dependent sensitivity variations as contributing sources of systematic error for our photometry. Since our science targets, Proxima Cen and Barnard's Star, are bright, the effect of Zodiacal Light is at most 0.001 magnitude.
Since we observe these stars only in the pickle center, spatially-induced variations
are reduced to about 0.001 magnitude, our claimed per-observation precision at $V \sim 11$. A weighted average of the temporal response of five stars in three fields provides a very precise flat field whose
slope error could introduce at most 0.001 magnitude systematic error over 1000 days. (Since we are doing only differential photometry we ignore the zero point error in the flat field.) Combining these sources of error yields a per-observation precision of 0.002 magnitude.

\section{Photometric Results}  \label{Photometric Results}
We  apply the flat field  (equation~\ref{FFeqn}) to the Appendix 1.1 and
Appendix 1.2 S values and plot (Proxima Cen, Figure~\ref{fig-9}; Barnard's
Star, Figure~\ref{fig-10}) the resulting intensities as
function of modified Julian Date (JD - 2440000). Our coverage in time is
not uniform for either target. There are extended gaps in coverage, some
due to the {\it HST}
solar constraint (no observations permitted closer than $\pm 50 \arcdeg$ to
the Sun). The largest gap, in 1994 for Proxima Cen, was due to an
awkward transition from Guaranteed Time Observations to Guest Observer
status and a hiatus due to
suspected equipment problems.

\subsection{Trends in Brightness}
For Proxima Cen the solid line in Figure~\ref{fig-9} indicates an overall trend of increasing brightness with time.  In units of normalized intensity the rate of change of brightness ($1.63 \pm 0.37 \times 10^{-5}$) is similar to that of the adopted flat field (equation~\ref{FFeqn}). For Barnard's Star (Figure~\ref{fig-10}) the slope of the upward trend in units of normalized intensity is $+0.91 \pm 0.18 \times 10^{-5}$, again, suspiciously similar in absolute value
to the adopted flat-field relation (equation~\ref{FFeqn}).  
Since seven stars from four different fields exhibit the sensitivity decrease discribed by the flat field, the 
Proxima and Barnard upward trends are unlikely to be a flat-field artifact.

A final caveat:  Proxima Cen and Barnard's Star are somewhat redder (Tables~\ref{tbl-1}
~and ~\ref{tbl-2}) than GJ 623 and GJ 748. If the sensitivity loss varies with wavelength (e.g., more sensitivity loss for blue than for red wavelengths),
it would have to be a very steeply dependent function, showing no effect 
at $B-V = +1.5$.

\subsection{Proxima Cen}

\begin{figure}
%\epsscale{1.0}
\plotone{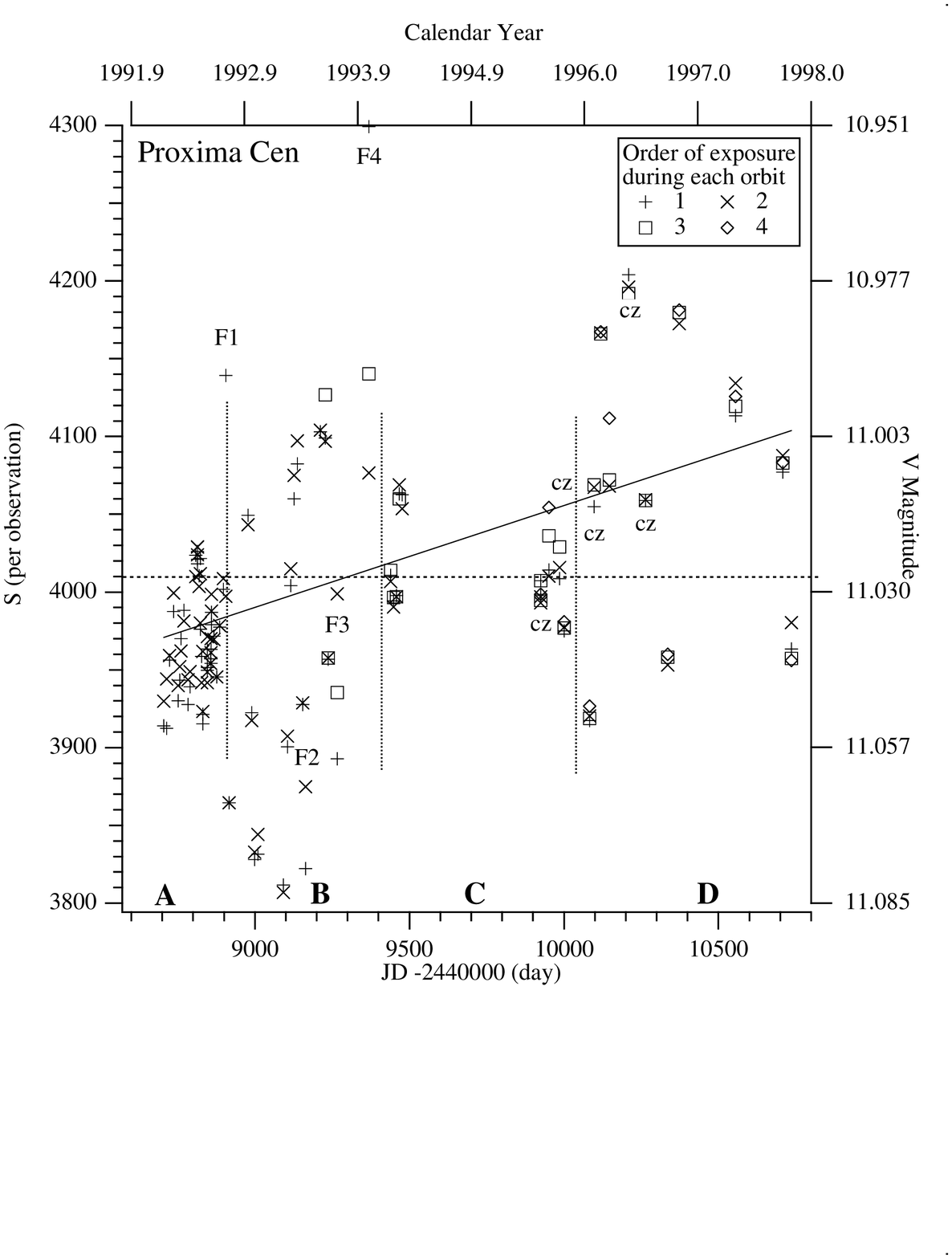}
\caption{Photometry of Proxima Cen. Each orbit contains 2, 3, or 4
exposures (Appendix 1.1).  Exposures perturbed by flares are marked F1 -
F4. Exposures acquired during CVZ orbits are labeled 'cz'.  Error bars are
about the size of the plotted symbols.  Four segments and two
distinct behavior modes are identified, A - D. Trend line is discussed in text.} \label{fig-9}
\end{figure}

The  flat-fielded photometry for each  exposure in each orbit appears
in Figure~\ref{fig-9}. The period and amplitude variations evident in
Figure~\ref{fig-9} will be discussed
in section 4. 
Our total time on target,
obtained by summing the exposure times in Appendix A.1, was 6\fh6. Proxima
Cen is a flare star (V654 Cen) and
these data contain exposures `contaminated' by stellar flares (marked F1 -
F4 in Figure~\ref{fig-9}). We identified these events by inspecting the
40Hz photometric data stream for each observation. An example of flare
contamination (including a light curve) can be found in
\cite{Ben93}, which discusses a  slow, relatively faint  ($\Delta V <
-0.10$), and multipeaked
flare on mJD 8906 (F1 in Figure~\ref{fig-9}).    An explosive flare on mJD
9266 ($\Delta V \sim 0 \fm 6$ in one second, F3 in Figure~\ref{fig-9})
produced
astrometric changes in Proxima Cen, analyzed in detail by  \cite{Ben97b}.
This spectacular event provided the
motivation for the subsequent CVZ observations (cz in
Figure~\ref{fig-9}),  each permitting 30 minutes of monitoring for flares.
The F4 event at mJD 9368 had a relatively small amplitude ($\Delta V \sim
-0.13$), but lasted throughout
the entire 130$^{s}$ exposure, hence its large effect on the exposure.
Walker (1981) predicts a  flare with intensity similar to
F3 once
every 31 hours. Thus, while disappointing, it is not surprising that we
captured none as bright as the
F3 event in our additional 2.5 hours of CVZ on-target monitoring. It may be
significant that we saw any flares at all, since even the small amplitude
events
have only a 60\% chance of occurring during our total monitoring duration. We will
discuss this further in Section \ref{PCSPOTS}.

Individual observations secured within an orbit and not affected by flaring
exhibit
an internal consistency at the 0.002 magnitude level.

\subsection{Barnard's Star} \label{Warm}
\begin{figure}
\plotone{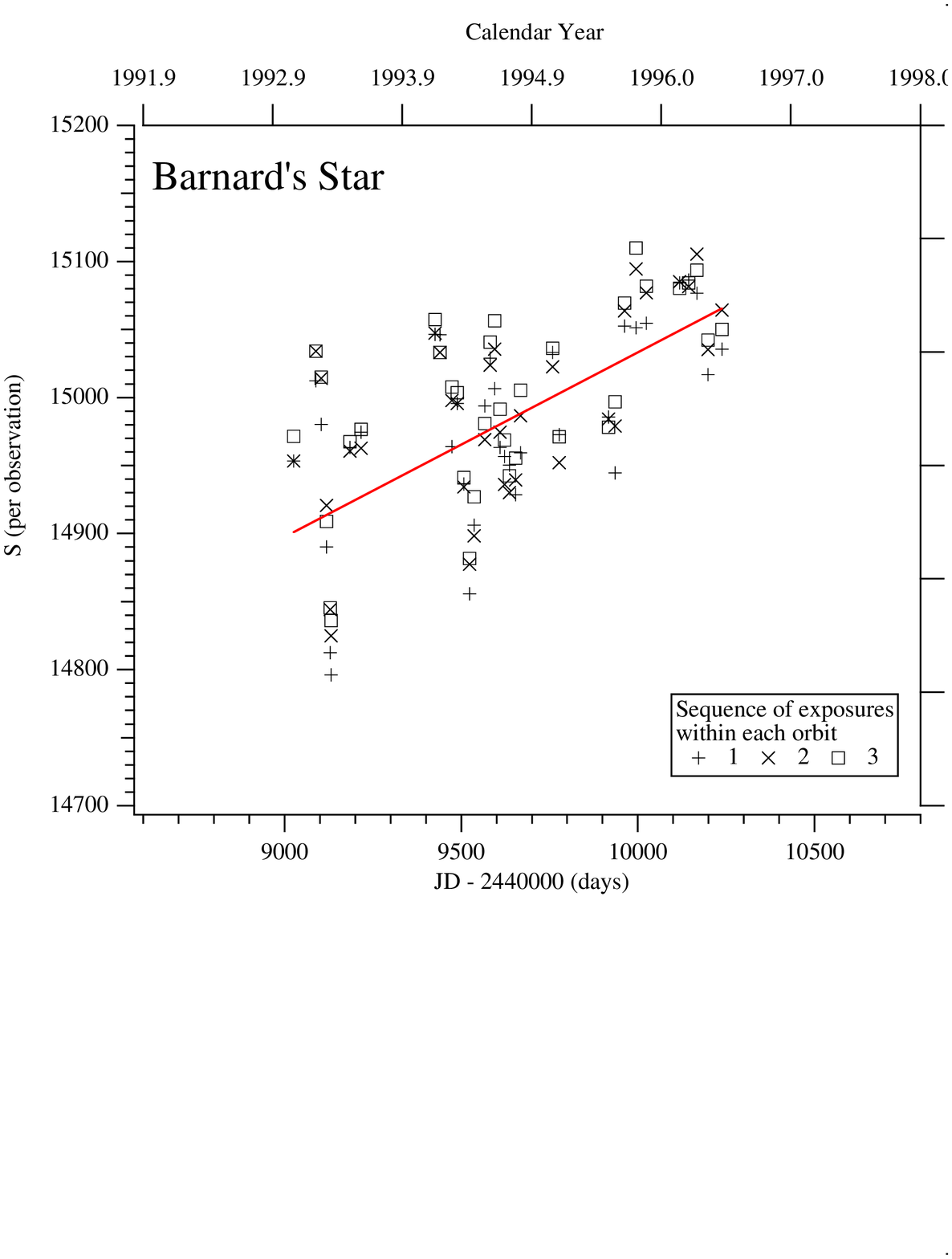}
\caption{Photometry of Barnard's Star. Each symbol represents the average
of an exposure.  Each orbit contains 3 exposures (Appendix 1.2). No flares
were detected. Error bars are smaller than the plotted symbols.  The time
scale is identical to Figure~\ref{fig-9}. Trend line is discussed in text. } \label{fig-10}
\end{figure}

The flat-fielded photometry for each  of the three Barnard's Star exposures
acquired within each orbit appears in
Figure~\ref{fig-10}. Note that the time scale is exactly that used for
Figure~\ref{fig-9} to facilitate comparison. Again, note that those
observations secured within an orbit exhibit
an internal consistency at the 0.002 magnitude level.  We find variations
within each orbit,
but no obvious flaring activity in the Barnard's Star
results.  Possible period and amplitude variations in the Barnard's Star
data will be discussed in
 section \ref{BSanal}.

The scatter within each orbit in Figure~\ref{fig-10} is somewhat larger than
the previously determined (\cite{Ben93})
0.001 magnitude measurement precision. In particular we inspected the
observations on mJD 9935 and 9994 and found only a slight upward slope
during the first observation on each date. Since the majority of first
observations within each orbit are lower, this intra-orbit scatter is most
likely an instrumental effect, amounting to about 0.001 magnitude. The
first observation low bias is sometimes seen in the Proxima Cen data
(Figure~\ref{fig-9}). Leaving all first observations uncorrected will only
slightly enlarge the formal errors for
our per-orbit means.

\section{Analysis}
For subsequent analyses of Proxima Cen, we removed
the flare contributions by subjecting the per-orbit average to
a pruning process. All exposures obtained during each orbit are presented
in Figure~\ref{fig-9}. If one exposure differs by more than $2.5-\sigma$
from the mean for that orbit,
it is removed and the mean recalculated. This process results in 71 normal
points with associated dispersions (the standard deviations calculated for
2, 3, or 4 exposures in each orbit) for Proxima Cen. No
exposures were removed from the Barnard's Star series, since no intra-orbit
points (shown in Figure~\ref{fig-10}) violated the
$2.5-\sigma$ criterion.
   The resulting per-orbit average S values are presented as direct light
curves
in Figure~\ref{fig-13} (Proxima Cen) and Figure~\ref{fig-16} (Barnard's Star).
Forming these normal points provides per-orbit photometric precision 
better than 0.002 magnitude
 for the following analyses.

\subsection{Lomb-Scargle Periodograms}
From Figures~\ref{fig-9} and~\ref{fig-10} we suspect that there are
periodic variations in
both the Proxima Cen and Barnard photometry. To obtain a preliminary
identification of these
periodicities we produce  Lomb-Scargle Periodograms  (\cite{Pre92}) from
the per-orbit normal points presented in Figure~\ref{fig-13} (Proxima Cen)
and Figure~\ref{fig-16} (Barnard's Star).

The most statistically significant period in the Proxima Cen periodogram
(Figure~\ref{fig-11}) is at $P \sim
83^{d}$, with a false-positive probability less than 0.1\%.
The very small peak at $P\sim42^{d}$ indicates the relative strengths of
the period derived from the first 212 days (\cite{Ben93}) relative to the
higher amplitude P$\sim83^{d}$ period. The false-positive probability for
the period derived in that paper from only the first 212 days was $\sim $1
\%. Since the periodogram provides no results for very short periods, we
have some
concern that we are undersampling a more rapid variation. We can rule out a
range of periods $2^{d} < P <20^{d}$  from
detailed inspection of clusters of data near mJD 8840
(Figure~\ref{fig-13}), where we had a series of closely-spaced (in time)
observations  (see Appendix 1.1).

The periodogram for Barnard's Star
 is shown in Figure~\ref{fig-12}. The strongest peak(at $P \sim 130^{d}$)
has a 10\% false-positive probability. We have much less compelling evidence of variability for
Barnard's Star than for Proxima Cen.

\subsection{Light Curves}
\subsubsection{Proxima Cen Light Curve} \label{PCanal}
\begin{figure}
%\epsscale{1.0}
\plotone{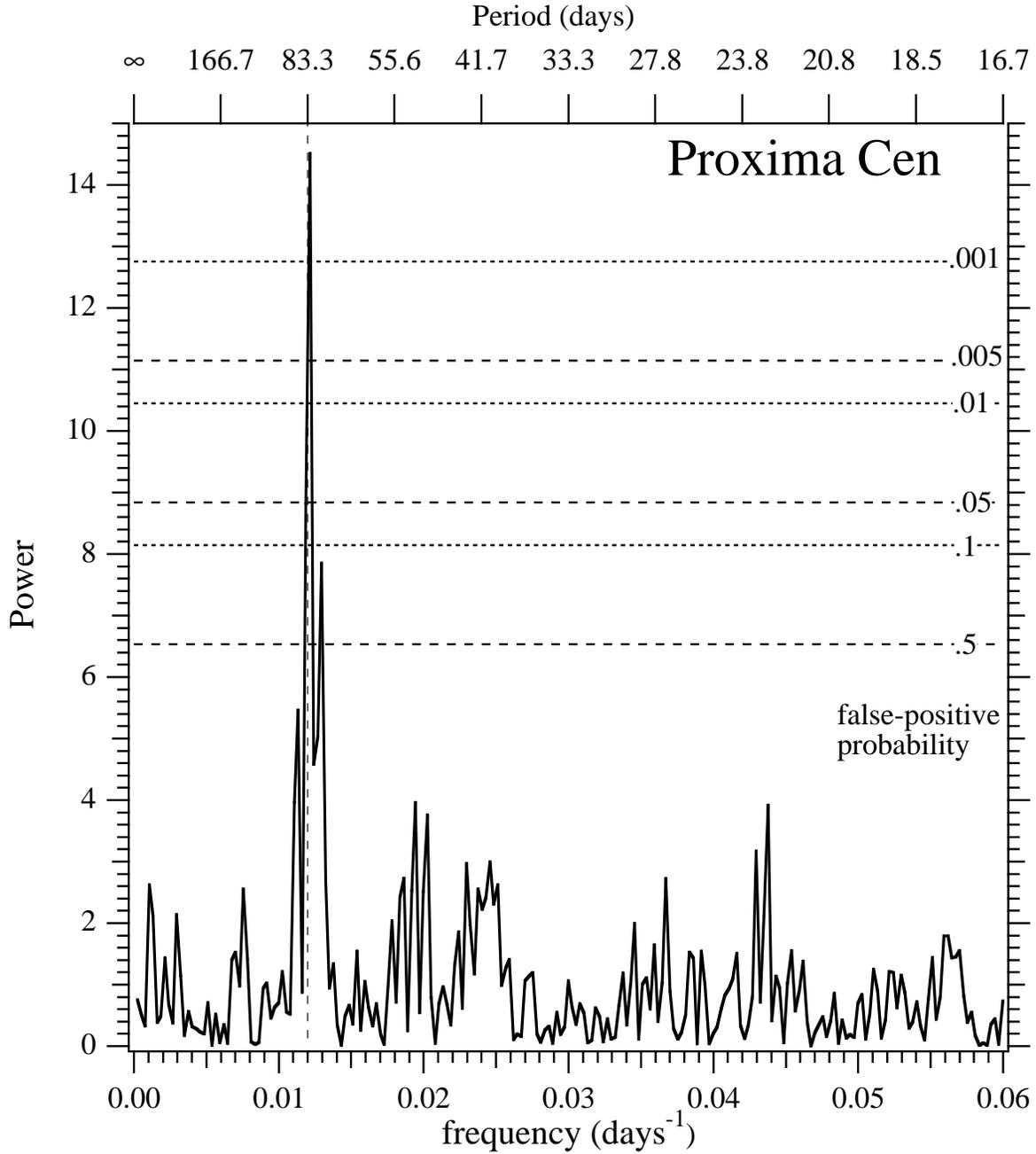}
\caption{ Periodogram from 71 normal points (average for
each orbit) with flares removed, flat-fielded with equation~\ref{FFeqn}.
The periodogram has most significant peak at $P\sim 83$ days,
with less than a 0.1\% false-positive probability. } \label{fig-11}
\end{figure}

\begin{figure}
\epsscale{1.0}
\plotone{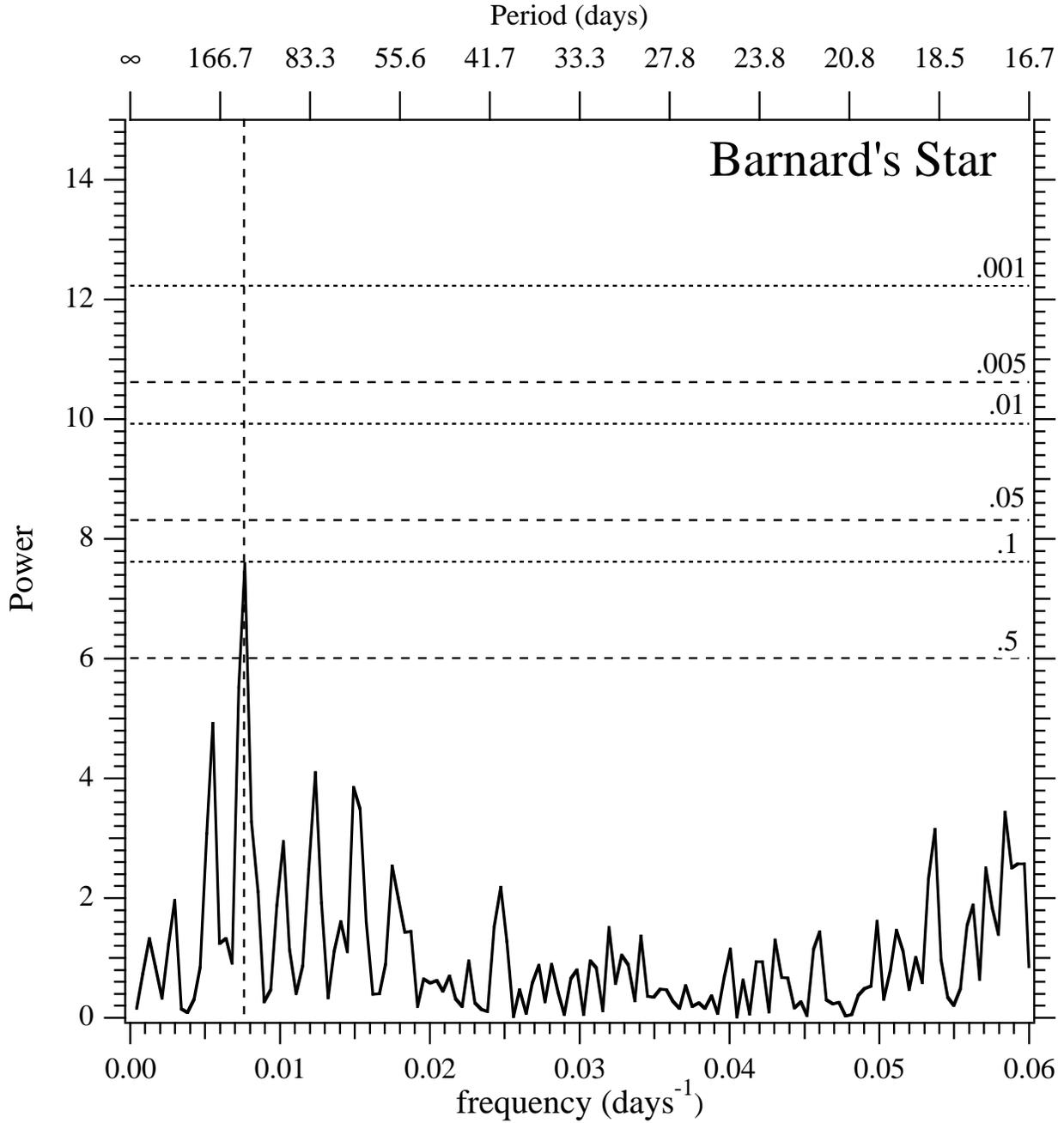}
\caption{Periodogram from 35 normal points (average for
each orbit), flat-fielded with equation~\ref{FFeqn}.
Periodogram has most significant peak at $P=130.4^{d}$,
with a 10\% false-positive probability.} \label{fig-12}
\end{figure}

Given strong support for a periodic variation (periodogram,
Figure~\ref{fig-11}) and for an overall trend in the brightness
(Figure~\ref{fig-9}), we  model the per-orbit average variations seen in
the direct light curve (Figure~\ref{fig-13}) with a sin function and trend
\begin{equation}
I = I_o + I't +A\sin( (\frac{2\pi}{P})t+\phi) ,\label{fiteqn}
\end{equation}
. 

To reconcile the earlier results (\cite{Ben93})
with the newer data, we first attempted to model the entire light curve
with only two distinct segments, grouping segments B, C, and D together.
From the earliest data (segment A) the Proxima Cen photometric variations
are characterized by a shorter period and smaller amplitude.  Later
data are best fit with a longer period and larger amplitude variation, as
evidenced by the periodogram (Figure~\ref{fig-11}). Parameters for these
fits are listed A and BCD in Table~\ref{tbl-4} (lines 1 and 2). We  find
$P_{BCD}/P_{A} = 1.97 \pm 0.04$.

Noting very large residuals for segment C, we next explored the possibility
that Proxima Cen repeats a low to high amplitude cycle by fitting  the four
segments (A - D, Figure~\ref{fig-13}) with the same model
(equation~\ref{fiteqn}). The parameters for these fits are presented in
Table~\ref{tbl-4} (lines 1 and 3 - 5). Within the errors, $P_{A} = P_{C}$
and
$P_{B} = P_{D}$, with  $P_{D}/P_{C} = 1.99 \pm 0.02$.
A reduction in the number of degrees of freedom by 17\% (fitting 71 data
points with twenty, rather
than ten parameters), reduced the residuals by $\sim$ 30\%. This relative
improvement is some support for alternating high and low amplitude states.
It
is also evident that segments A and C have very nearly half the period
of segments B and D.

Figure~\ref{fig-14} contains phased light curves for the four  segments
labeled in Figure~\ref{fig-13}. In the top panel we show that the phase
shift required to align the two long-period segments is small
($\Delta\phi =
+0.11$) for the period $P = 83.5^{d}$ suggested by the periodogram
(Figure~\ref{fig-11}).
We have shifted  segment D down by $\Delta$S = -104.4 counts. The bottom
panel of Figure~\ref{fig-14} shows a phased light curve for the two shorter
period segments (A and C), phased also to $P = 83.5^{d}$. Shifts in phase
and intensity to achieve alignment  are indicated in the figure. The clean
double sin wave also demonstrates
that the low-amplitude segments, A and C, have half the period of the
higher-amplitude, longer-period segments, B and D.

Finally, Figure~\ref{fig-15} is used to demonstrate that the same low-amplitude,
short-period variations seen in segments A and C may also be present in
segments B and D. We fit a sin wave to the phased B and D light curve in
the lower panel of Figure~\ref{fig-15}, constraining the period to one cycle.
The top panel of Figure~\ref{fig-15} shows the residuals to that fit. These
residuals are then
fit with a sin wave with the period constrained to one-half cycle. Comparing
with the bottom panel of Figure~\ref{fig-14}, we find  a similar double sin
wave, nearly identical
phase, and an amplitude ($A = 25 \pm 8$) close to that reported for
segments A and C in Table~\ref{tbl-5}.

\subsubsection{Barnard's Star Light Curve} \label{BSanal}
We turn now to the per-orbit average photometry of Barnard's Star.
Figure~\ref{fig-16} (bottom) contains the per-orbit average direct light
curve. The error bars are
$1-\sigma$, obtained from the dispersion of the three observations within
each orbit on each date (Figure~\ref{fig-10}).  Residuals
to a linear trend are presented
in the top panel of Figure~\ref{fig-16}. The sin wave fit to these residuals
was constrained to have the most significant period detected in the
Figure~\ref{fig-12} periodogram, $P = 130\fd4$.
Figure~\ref{fig-17} contains  a light curve for the trend-corrected Barnard's
Star photometry of Figure~\ref{fig-16} (top), phased to $P = 130\fd4$. The
phased light curve is far less clean than for Proxima Cen. The periodogram
and Figures ~\ref{fig-16} and ~\ref{fig-17} provide only weak evidence for
periodic variation, primarily due to the poor sampling.

\section{Discussion of Photometric Results}
Instruments can impress spurious periodicities on data (\cite{Kri91}).
It is comforting that we find for all segments of either data set that
$P_{Barn} \ne P_{Prox}$.

Stars have local imperfections in their atmospheres (e.g., the Sun,
\cite{Zir88}). Stars
other than the Sun have been shown to be spotted, photometrically (dwarf M
stars; \cite{Kro52}) and spectroscopically (e.g., \cite{Hat93};
\cite{Nef95}). Other M stars have been
shown to have spots, both dark (\cite{Bou95}) and bright ($\alpha$ Ori,
\cite{Gil96}).

A spot on a rotating star is a model rich in degrees of freedom. Spots can
be bright (plages) or dark (see \cite{Pet92} for a discussion of the choice
between dark spots on a bright background or bright spots on a dark
background).
Spots can wax and wane in size, driving the mean brightness level of a star
up or down (\cite{Vrb88}). Spots can migrate in latitude, which, when
coupled with presumed differential rotation, can change the phasing and
perceived rotation period. Spots are thought to migrate up or down
(relative to the star center) within the magnetosphere (\cite{Cam93}),
inducing perceived period
changes. In the following sections we shall interpret the variations seen
in Figures~\ref{fig-13}~and~\ref{fig-16}  as rotational modulation of
spots or plages.

\subsection{Spots on Proxima Cen} \label{PCSPOTS}

\begin{figure}
\epsscale{1.0}
\plotone{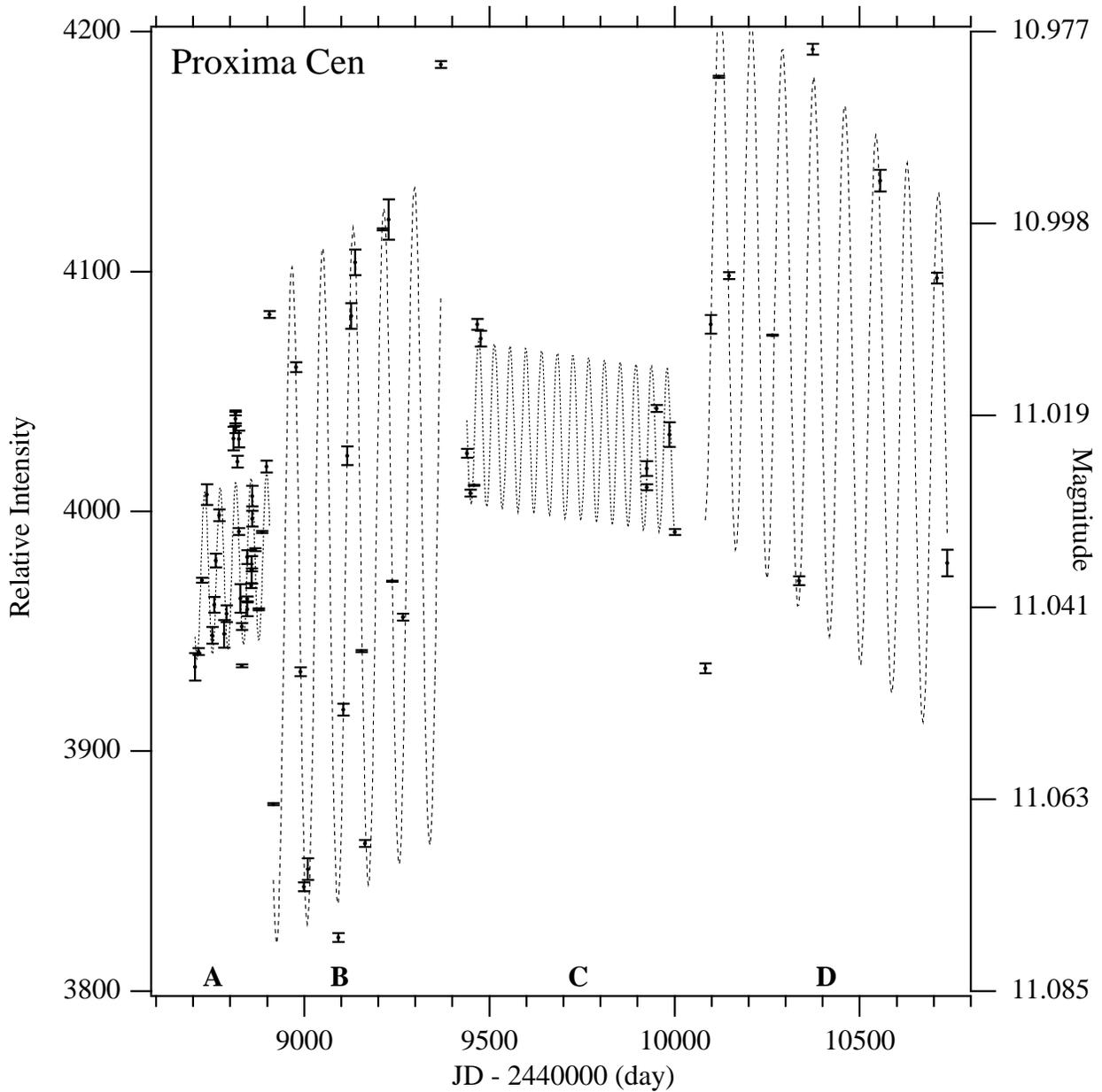}
\caption{Direct light curve. Each symbol represents the average of from two
to four exposures per orbit, with flares removed.   Error bars are
$1-\sigma$. See Table~\ref{tbl-5},  lines 1, 3, 4, and 5 for the results of
fitting a sin wave and trend (equation \ref{fiteqn}) to each segment. }
\label{fig-13}
\end{figure}

 If we assume a fundamental rotation period  $P = 83.5^{d}$, then
variations in the amplitude (Figure~\ref{fig-13}) could be due to
spot/plage changes. With the sparse set of single-color photometric data
presented in Figure~\ref{fig-13} we
have made no effort to quantitatively model spots (c.f. \cite{Nef96}). The
period and amplitude changes can be qualitatively
 modeled using plages and spots,  but require the disappearance of a
feature or a major change in
feature size or temperature in less than one rotation period (e.g., the A to B
 segment transition seen in Figure~\ref{fig-13}).

\begin{figure}
\epsscale{.6}
\plotone{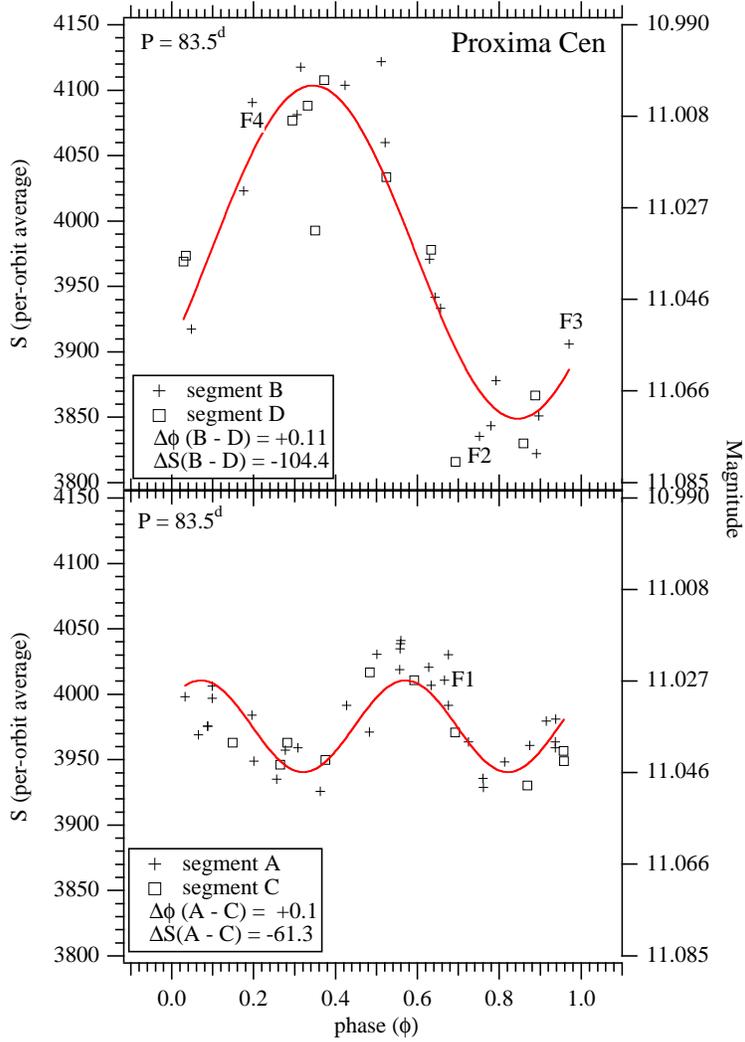}
\caption{Phased light curves for Proxima Cen. Top: the two long-period segments
(B and D, Figure~\ref{fig-13}). Sin wave fit has a period constrained to
one cycle.  Bottom: short-period segments (A and C) are
phased to the longer period and show a double sin wave. The fit is
constrained to
have a period of one-half cycle.   Error bars are about the
size of the symbols. The observed flares are labeled F1 - F4.} \label{fig-14}
\end{figure}

\begin{figure}
%\epsscale{.6}
\plotone{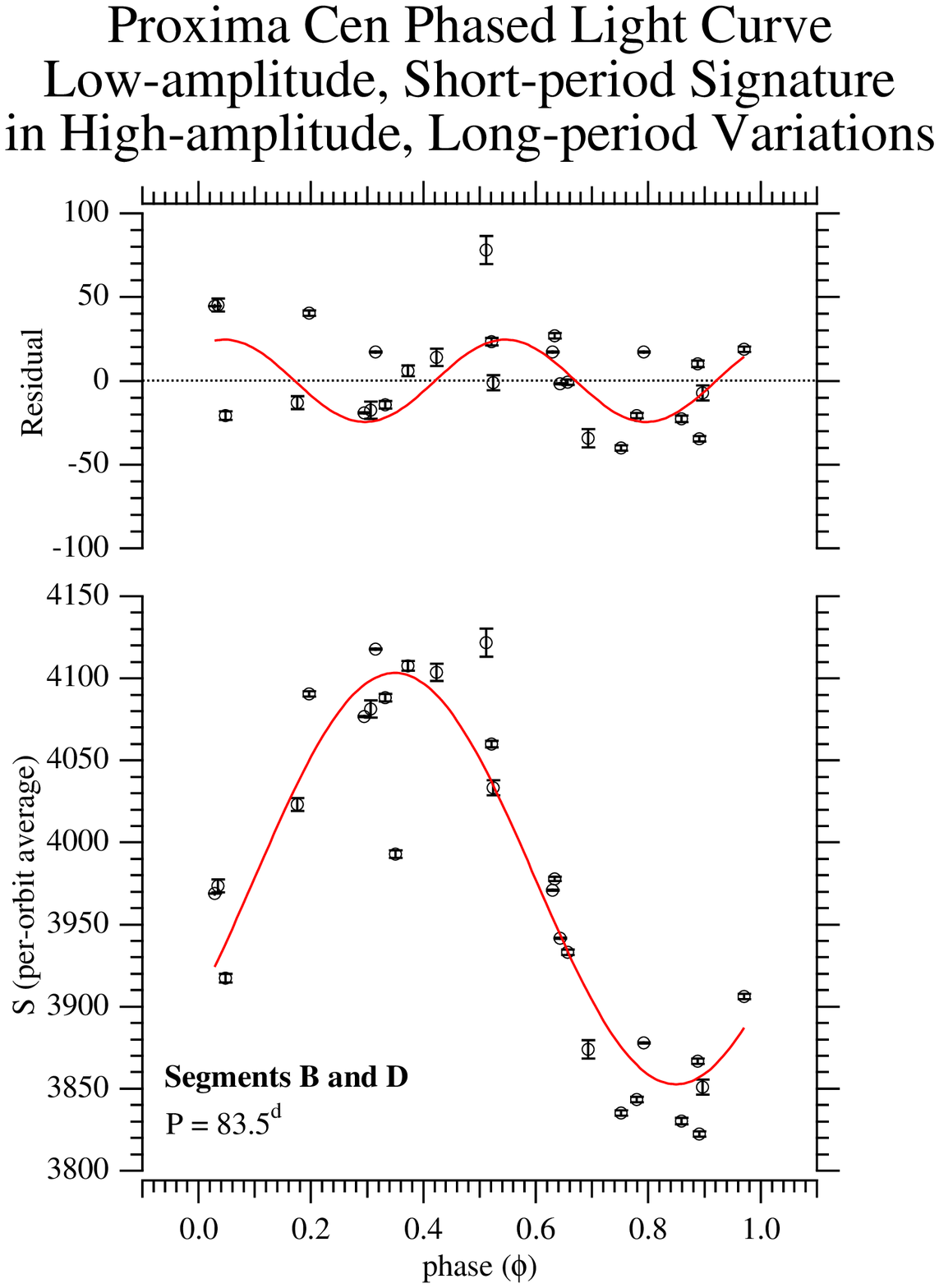}
\caption{Phased light curve for Proxima Cen. Bottom: the two long-period
segments
(B and D, Figure ~\ref{fig-13}), phased to $P = 83\fd5$ and fit with a sin
wave having a period of one cycle.  Top: the residuals to the sin
wave fit. These show a double sin wave pattern nearly identical to the two
short-period segments, A and C (Figures~\ref{fig-13} and~\ref{fig-14}),
suggesting that the low-amplitude, short-period signature persists during
segments
B and D.} \label{fig-15}
\end{figure}

\begin{figure}
%\epsscale{1.0}
\plotone{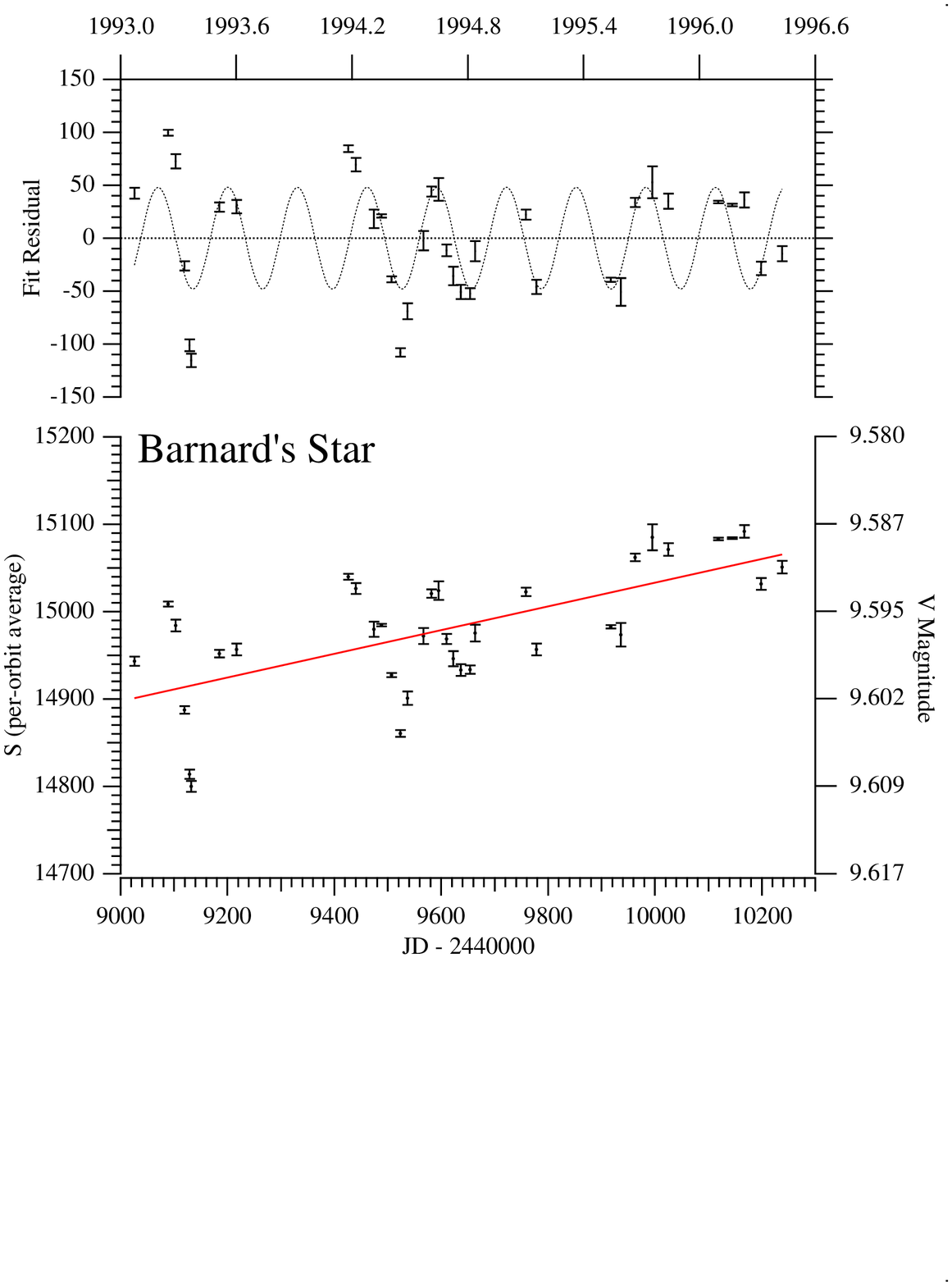}
\caption{Per-orbit average direct light curve. Each symbol represents the
average of three exposures.  Error bars are $1-\sigma$. Bottom:  fit with a
trend line.
Top: residuals to trend line,  fit with a constant-amplitude sin wave with
constrained $P=130\fd4$. } \label{fig-16}
\end{figure}

\begin{figure}
%\epsscale{1.0}
\plotone{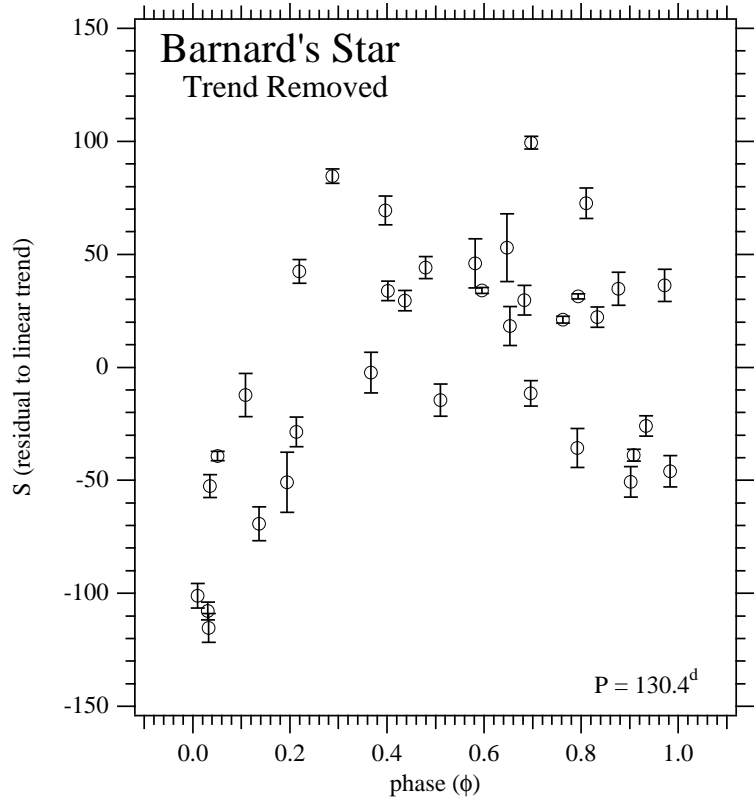}
\caption{Trend-corrected light curve for Barnard's Star, phased to
$P=130\fd4$. The full-amplitude of the variation is $\sim0.01$ magnitude.
Error bars are $1-\sigma$. } \label{fig-17}
\end{figure}

Segments B and D (Figure~\ref{fig-14}, top) require a single large or
darker spot that reduces the average brightness of Proxima Cen by $\Delta V
\sim 0.03$. This feature was not present in segment A and disappeared
during segment C. To phase  segments B and D we applied a shift of $\Delta
\phi = 0.1$ radians. Thus, the spot site lagged behind the fundamental
rotation period by
about $5\arcdeg$ over the end of B to start of D time separation
of $\sim 700^{d}$. Whether due to latitude migration coupled with
differential rotation or from changes of height within the magnetosphere is
unknown. If due
to differential rotation, then either the spot moved very little in latitude
or differential rotation on Proxima Cen is several orders of magnitude
less than the Sun (\cite{Zir88}).

Segments A and C exhibit smaller amplitude variations with a period almost
exactly half that found for segments B and D. These segments (see the
phased light curves in Figure~\ref{fig-14}, bottom and Figure~\ref{fig-15},
top) could be produced by two smaller spots spaced $180\arcdeg$ in
longitude, carried
around by the fundamental 83\fd5 rotation period, and
persisting through all segments A to D. These two spots produce a $\Delta V
\sim 0.01$. One of these small or less dark (warmer) spots lies at
nearly the same longitude as the prominent spot seen in segments B and D.
The other lies near the center of the brighter hemisphere in segments B and D.

From Figure~\ref{fig-13} we note that
the peaks in segments B and D were brighter.
In segment B the minima were deeper, implying a darker (cooler) spot. To
increase the amplitude of the maxima in segments B and D requires the
existence of
plages, or that the hemisphere not containing the single large spot became
brighter due to spot changes. One of the pair of spots, associated with the
brighter hemisphere
($\phi \sim 0.3$, Figure~\ref{fig-15}, top), does have a shallower minimum
than the other of the pair at $\phi = 0.8$. If these are the same spots
responsible for the variation in segments A and C, then the spot at $\phi
\sim 0.3$ did have a shallower minimum
in segments B and D (compare Figure~\ref{fig-14}, bottom and
Figure~\ref{fig-15}, top). However, that spot did not become less dark by
enough to account for the increased maxima seen in segments B and D. As a
consequence
we propose
plage activity to increase the segment B and D maxima.

Flaring activity seems more prevalent in
 segment B. As seen in Figure~\ref{fig-14}, three of four flares are
grouped near phase $\phi = 0.8$. Association with this deep minimum might
imply some connection of flaring activity with the largest or coolest star spot,
which is also at the same longitude as one of the two smaller spots seen
best in segments A and C. The remaining flare, F4, lies close to $\phi =
0.2$, the other spot of the
low-amplitude pair of spots.
Spot/flare association was previously noted in the M dwarf EV Lac,   shown
also to have longitude-dependent
flaring associated with a star spot site (\cite{Let97}). However, for EV Lac,
 flares were detected a year before the spots became easily detectable by
their system ($\Delta V \sim 0.1$) and, once spots formed, flare activity
abated. 

\subsection{A Small, Variable Spot on Barnard's Star}
	The periodogram (Figure~\ref{fig-12}) does not provide a clear
identification of a single period of variation. The trend-corrected
direct light  curve (Figure~\ref{fig-16}, top) has been fit
with a sin wave, constraining $P= 130\fd4$. The constant amplitude is
$\Delta V \sim 0 \fm 01$, about five times our formal photometric error.
Given the sparse coverage, it is speculative to interpret this light curve as showing rotational modulation of a single, small spot decreasing in size.

\subsection{Rotation Periods for Proxima Cen and Barnard's Star}
Rotation periods for Proxima Cen have been  predicted from chromospheric
activity levels by Doyle (1987), who obtained P=$51^{d} \pm
12^{d}$. Guinan \& Morgan (1996) measure a rotation period (P = $31 \fd 5
\pm 1\fd 5$) from IUE observations of strong \ion{Mg}{2}
h+k emission at 280nm.  We find no support for either rotation period
in our periodogram (Figure~\ref{fig-11}) or light curves,  direct
(Figures~\ref{fig-13}) or phased (Figure~\ref{fig-14}). We do note that the
variation due to the spot pair produces a period between the Doyle prediction
and Guinan \& Morgan measurement.

The observed variations for Proxima Cen and Barnard's Star, if interpreted
as rotationally modulated spots, yield rotation periods far longer than for
other M stars. For example, \cite{Bou95} find $4^{d} < P < 8^{d}$ for a
sample of young, early-M T Tauri stars. Magnetic braking is postulated to
slow rotation over time (\cite{Cam93}). The inferred rotation period for
Proxima Cen is
consistent with old age. This age  may be 4 - 4.5 By, if Proxima Cen is
coeval with $\alpha$ Cen (\cite{Dem86}).
A relatively older age for Barnard's Star  can be surmised from lower than
solar metallicity (\cite{Giz97}) and higher than solar space velocity
(\cite{Egg96}),
both consistent with a longer rotation period, if one accepts the reality of the variation.

\subsection{Shorter Time-scale Variations}
The level of internal per-orbit precision for these photometric data is
near 0.002 magnitudes. Hence, the dispersion about the phased light
curves for
Proxima Cen (Figure~\ref{fig-14}) is likely intrinsic to the stars. Two possibilities are miniflaring  and
the creation and destruction of small star spots and plages.  That either
phenomenon must
have  a duration longer than hours, at least for Proxima Cen, is suggested by
the segment A phased light curve (Figure~\ref{fig-14}, bottom)
and a detailed light curve for segment A (see~\cite{Ben93}, Fig. 3).
Segment  A contains four pairs of back-to-back orbits and one set of three
contiguous orbits (on mJD 8845).  In each case the time on target coverage
is over 90 minutes. For most of these contiguous orbits differences are
within two standard deviations and not statistically significant. Since 'flare'
implies a relatively short duration, miniflaring cannot be the cause
of the scatter.

\subsection{Activity Cycles}
The $\sim 1100^{d}$ cycle of alternating high and low
amplitude (see Figure~\ref{fig-13}) is suggestive of an activity cycle for
Proxima Cen.
However, the gap in our coverage in segment C  weakens any
claims that can be made relative to the timing of this cycle. Comparing their
1995 IUE data with earlier archival data, \cite{Gui96} propose an activity
cycle that was in a low-state in 1995, agreeing with our identification
of segment C representing a low-state (Figure~\ref{fig-13}).

\section{Conclusions}

1.~For FGS 3 photometry we have identified four sources of systematic error:
background contamination (primarily Zodiacal Light); spatial flat field variations (significant only for target positions $r > 20\arcsec$ from pickle center); temporal sensitivity changes (calibrated to a level introducing a
0.001 magnitude differential run-out error in 1000 days); and a possible warm-up effect
(see section~\ref{Warm}). 

2. ~Two to four short ($t \le 100^{s}$) observations with FGS 3 during one
orbit yield 2 milli-magnitudes precision photometry, provided the targets are
bright ($V \le 11.0$) and restricted to the central 20\arcsec ~of the pickle.

3. ~Proxima Cen exhibits four distinct segments with two distinct behavior
modes: short period, low amplitude and long period, large amplitude. These
variations are consistent with a fundamental rotation period,
$P = 83\fd5$ and three darker spots.
Two of the spots are either very small or very low-contrast. They are spaced by
180\arcdeg ~and persist throughout our temporal coverage, over 24 rotations.
A single more prominent spot (either large or high-contrast)
formed in less than one rotation period, persisted through four rotations, then
disappeared. A spot reappeared within 5\arcdeg ~of this same longitude five
rotations later. The hemisphere opposite this spot brightened each time the
spot formed.

It is intruiguing that active longitudes of spot formation separated by 180\arcdeg ~are observed
in chromospherically active stars with close stellar companions (\cite{Hen95}). If the photometric behavior of Proxima Cen is indicative of a synchronously rotating companion, its mass is less than that of Jupiter (\cite{Ben97a}).

4. ~We interpret the four distinct segments with two distinct behavior
modes  seen in the Proxima Cen
photometry as an activity cycle with a period $\sim 1100^{d}$. Most of the flare
activity occurred in the long period, high amplitude variation segment B.
In the phased lightcurve three of the four detected flares are near
the deepest minimum.

5.~The scatter in the Proxima Cen phased light curve is far larger than our photometric
precision. This scatter could be caused by  the
formation and  dissolution of small spots or plages within one rotation period.

6. ~ We find brightness variations five times our
formal photometric precision for Barnard's Star. Unfortunately,
the sparse coverage of the possible variation renders it a marginal detection. 
We conclude that Barnard's Star shows very weak evidence for periodicity on a
timescale of approximately 130 days

To confirm the spots and the inferred rotation periods will require observations of color changes (e.g. \cite{Vrb88}) and additional
spectroscopic observations of lines sensitive to presence or absence of
star spots. Extended-duration milli-magnitude V band photometry from the
ground, while difficult (\cite{Gil93}), could
probe the activity cycle periodicity of Proxima Cen. 
Future tests could include {\it Space Interferometry Mission}
(\cite{Sha95}) observations with several microarcsec astrometric precision.
If spots and plages exist on these stars, we can expect
easily detectable star position
shifts as activity sites vary. Such observations will provide detailed maps
of spot and plage location. Extended temporal monitoring will provide
evolutionary details.

\acknowledgments

Support for this work was provided by NASA through grants GTO NAG5-1603 and
GO-06036.01-94A from the Space Telescope Science Institute, which is operated
by the Association of Universities for Research in Astronomy, Inc., under
NASA contract NAS5-26555. We thank Bill Spiesman and Artie Hatzes for
discussions and draft paper reviews and Melody Brayton for paper
preparation assistance. Denise Taylor provided crucial scheduling
assistance at the Space Telescope Science Institute. Suggestions by an anonymous referee
produced substantial improvements to the original draft.
\section*{Appendix 1}
Appendix 1 contains the observation logs and measured average S values for
Proxima Cen and Barnard's Star.
\clearpage
%\placefigure{fig1}
%\placetable{tbl-3}

% Now comes the reference list.  In this document, we used \cite to call
% out citations, so we must use \bibitem in the reference list, which
% means we use the LaTeX thebibliography environment.  Please note that
% \begin{thebibliography} is followed by a null argument.  If you forget
% this, mayhem ensues, and LaTeX will say "Perhaps a missing item?" when
% you run it.  Do not call us, do not send mail when this happens.  Put
% the silly {} after the \begin{thebibliography}.
%
% Each reference has a \bibitem command to define the citation format
% to be placed in the text (in []) and the symbolic tag used for
% cross referencing (in {}).
%
% See sample1.tex, or the AASTeX guide, for an alternative to the \cite-
% \bibitem command.

\clearpage
\begin{center}
\begin{deluxetable}{lll}
\tablecaption{Proxima Cen  \label{tbl-1}}
\tablewidth{0in}
\tablehead{\colhead{Parameter} &  \colhead{Value}&
\colhead{Reference}}
\startdata
aliases&$\alpha$ Cen C, GJ 551, V645 Cen& \nl
$M_{V}$ & $15.45\pm0.1$ &   \nl
\bv & 1.94 \\
Sp.T. & M5Ve &    \cite{Gli91}\nl
$M_{\rm Prox}$ & $0.11M_{\sun} $  &\cite{Kir94}\nl
$L_{\rm Prox}$ & $0.001L_{\sun}$  &\cite{Lie87}\nl
$R_{\rm Prox}$ &  $0.15R_{\sun}$& \cite{Pan93}
\enddata
\end{deluxetable}
\end{center}

\clearpage

\begin{center}
\begin{deluxetable}{lll}
\tablewidth{0in}
\tablecaption{Barnard's Star  \label{tbl-2}}
\tablehead{\colhead{Parameter} &  \colhead{Value}&
\colhead{Reference}}
\startdata
aliases & GJ 699, G 140-24,  LHS 57& \nl
$M_{V}$ & $13.2\pm0.1$ &   \cite{Gli91}\nl
\bv & 1.73\nl
Sp.T. & M4Ve &   \cite{Kir94}\\
$M_{\rm Barn}$ & $0.17M_{\sun} $  &\cite{Hen93}\nl
$L_{\rm Barn}$ & $0.0046L_{\sun}$  &\cite{Hen93}\nl
$R_{\rm Barn}$ &  $0.17R_{\sun}$&
\enddata
\end{deluxetable}
\end{center}

\clearpage

\begin{center}
\begin{deluxetable}{lccc}
\tablewidth{0in}
\tablecaption{Flat Field Modeling - M35 Stars \label{tbl-3}}
  \tablehead{\colhead{Parameter} &  \colhead{S9\_79}&
\colhead{S11\_68}&  \colhead{S9\_47}}
\startdata
linear trend \nl
I$_o (counts)$    &  1.134  &1.107 &1.129\nl
   &   $\pm  0.008$ & 0.023 & 0.015\nl
I' (counts $d^{-1}$) &    $-1.36$E-05   &    $-1.11$E-05   &    $-1.31$E-05\nl
  &      0.10E-05   &     0.24E-05   &     0.10E-05\nl
Variation from $180\deg$ flip \nl
A (counts)   &   0.0006 & -0.0065 & $0.0039$\nl
   &    0.0006  &0.0003&  0.0004\nl
P (days)   &   365.25 & 365.25 & 365.25\nl
    &    0    &   0  &     0\nl
$\phi$ (radians)  &   281.56& 281.71& 282.00\nl
  &    1.03 &  0.17  &  0.14\nl
\enddata
\end{deluxetable}
\end{center}
\clearpage
\begin{center}
\begin{deluxetable}{lc}
\tablewidth{0in}
\tablecaption{Flat Field Modeling - M Dwarf Stars \label{tbl-4}}
  \tablehead{\colhead{Parameter} &  \colhead{GJ 623 and GJ 748}}
\startdata
I$_o (counts)$    &  1.133  \nl
   &   $\pm   0.0132$ \nl
I' (counts $d^{-1}$) &    $-1.30$E-05   \nl
  &      0.13E-05        \nl
\enddata
\end{deluxetable}
\end{center}
\clearpage

\begin{deluxetable}{lrrlrlrlrlr}
\tablewidth{0in}
\tablenum{5}
\tablecaption{Proxima Cen Light Curve Parameters \label{tbl-5}}
\tablehead{& & \colhead{$I_o$}&& \colhead{$I'$}&& \colhead{A} && \colhead{P} &&
\colhead{$\phi$} \\ \cline{3-3}  \cline{5-5}  \cline{7-7} \cline{9-9}
\cline{11-11}
\colhead{Segment } & \colhead{N}  & \colhead{(counts)} & &
\colhead{(counts d$^{-1}$) } &&
\colhead{(counts) } & &  \colhead{(days)} & &\colhead{(radians)}}
\startdata
 A &     32 &    3560\phantom{00} & & 0.05\phantom{0000} & &
$34.2$\phantom{00} &
&       41.8\phantom{00} &&  $-1.6$\phantom{00} \nl
 &&    $\pm  710$\phantom{00}  &   & 0.08\phantom{0000} & &
6.0\phantom{00} & &
0.9\phantom{00} & &  2.7\phantom{00}\nl
BCD &  39 &    3200\phantom{00} & & 0.09\phantom{0000} && $-102$\phantom{00} &
&        82.5\phantom{00} & & -0.4\phantom{00} \nl
 &&      160\phantom{00} &&   0.02\phantom{0000} &&  12\phantom{00} &&
0.3\phantom{00} & &  2.4\phantom{00}\nl
B &     17&     3070\phantom{00} && 0.10\phantom{0000} &&
$-139$\phantom{00} & &
82.7\phantom{00} &&$ -2.8$\phantom{00} \nl
&&      580\phantom{00}&  & 0.06\phantom{0000} &&  9.5\phantom{00}  &&
0.7\phantom{00}  &&  0.7\phantom{00}
\nl
C       & 11    &
4250\phantom{00}&& $ -0.02$\phantom{0000} && $-34.5$\phantom{00} &&
42.4\phantom{00}&& $ -2.9$\phantom{00} \nl
&&            200\phantom{00} &&  0.02\phantom{0000} && 8.0\phantom{00} &&
0.2\phantom{00} && 1.3\phantom{00} \nl
D       & 11 &  5473\phantom{00} && $-0.14$\phantom{0000}&&
$-114.8$\phantom{00} && 84.3\phantom{00} && $ -1.5$\phantom{00} \nl
              &&        645\phantom{00} && 0.06\phantom{0000}  &&
16.3\phantom{00}  && 0.8\phantom{00} &&  2.2\phantom{00}
\enddata
\end{deluxetable}

\clearpage

% And finally, we must deal with the figures.  There are three figures
% associated with this manuscript; two figures are Encapsulated
% PostScript (EPS) files.  The third figure is a grey scale figure that does
% not exist in EPS form.
%
% Authors have three options for including figure information within a
% manuscript.  Not all the options may be acceptable by the target Journal - be
% sure to look at the appropriate submission instructions, electronic or
% otherwise.
%

\clearpage

\clearpage

\clearpage
\clearpage

\clearpage
\clearpage
\clearpage

\clearpage
\clearpage

\clearpage

\clearpage

\clearpage
\clearpage

\clearpage
\clearpage
\clearpage

\clearpage
\clearpage

% The \plotone and \plottwo commands scale the plot(s) in both dimensions
% so that the horizontal dimension fits in the body of the text.  The
% \plotfiddle command will override any automatic scaling, but often
% requires additional "fiddling" to get the plot to fit on the page.
% The \epsscale command allows the author to simply change the scaling
% of the plot in place, without the additional "fiddling" required by
% \plotfiddle.

%\begin{figure}
%\epsscale{.6}
%\plotone{sgi9259.eps}
%\caption{This is an example of a long figure caption that must be set as
%a paragraph.  The processor has to buffer the text of the
%caption, so it is good not to be too wordy, but that would make for
%poor communication as well. \label{fig3}}
%\end{figure}

% That's all, folks.
%
% The technique of segregating major semantic components of the document
% within "environments" is a very good one, but you as an author have to
% come up with a way of making sure each \begin{whatzit} has a corresponding
% \end{whatzit}.  If you miss one, LaTeX will probably complain a great
% deal during the composition of the document.  Occasionally, you get away
% with it right up to the \end{document}, in which case, you will see
% "\begin{whatzit} ended by \end{document}".

\end{document}

\begin{deluxetable}{rrllrrl}
\tablewidth{0in}
\tablenum{A1.1}
\tablecaption{Proxima Cen Log of Observations } 
\tablehead{& \colhead{Exposure} & \colhead{Average} & & &\colhead{Exposure}
& \colhead{Average}
\\
\colhead{mJD} &
\colhead{Time} &
\colhead{Intensity} &&\colhead{mJD} & \colhead{Time} & \colhead{Intensity}}
\startdata
8704.858&93.60\phantom{00} &3972.96 && 8858.714& 92.40\phantom{00} &4038.52\nl
8704.871& 66.45\phantom{00} &3988.01 && 8858.733& 64.35\phantom{00} &4049.20\nl
8713.824& 92.55\phantom{00} &3971.07 && 8866.820& 92.40\phantom{00} &4021.89\nl
8713.838& 66.45\phantom{00} &4001.50 && 8866.837& 79.35\phantom{00} &4019.69\nl
8723.807& 92.40\phantom{00} &4015.08 && 8876.226& 92.40\phantom{00} &3995.54\nl
8723.820& 66.45\phantom{00} &4016.87 && 8886.134& 92.25\phantom{00} &4027.40\nl
8736.396& 93.45\phantom{00} &4044.73 &&  8886.151& 79.35\phantom{00} &4027.74\nl
8736.410& 67.50\phantom{00} &4055.95 &&  8896.979& 92.40\phantom{00} &4051.47\nl
8751.390& 92.40\phantom{00} &3986.84 && 8896.996& 79.50\phantom{00} &4057.43\nl
8751.404& 66.45\phantom{00} &3995.15 && 8906.083& 92.55\phantom{00} &4192.68\nl
8756.509& 92.40\phantom{00} &3999.22 &&  8906.101& 79.35\phantom{00} &4045.50\nl
8756.524& 66.30\phantom{00} &4007.55 && 8916.583& 92.40\phantom{00} &3911.94\nl
8759.860& 92.40\phantom{00} &4026.46 &&  8916.597& 79.50\phantom{00} &3912.24\nl
8759.873& 66.45\phantom{00} &4016.65 && 8977.539& 92.40\phantom{00} &4095.24\nl
8769.707& 92.40\phantom{00} &4044.77 && 8988.275& 92.25\phantom{00} &3966.21\nl
8769.720& 66.45\phantom{00} &4036.54 && 8988.293& 79.35\phantom{00} &3960.82\nl
8783.702& 92.40\phantom{00} &3981.93 && 8999.116& 92.25\phantom{00} &3870.63\nl
8783.715& 66.60\phantom{00} &3997.98 && 8999.134& 79.50\phantom{00} &3874.66\nl
8790.130& 92.40\phantom{00} &3992.80 && 9008.892& 92.25\phantom{00} &3873.44\nl
8790.143& 67.35\phantom{00} &4002.20 &&  9008.909& 79.35\phantom{00} &3886.46\nl
8813.522 & 92.25\phantom{00} & 4072.92 && 9091.916& 92.25\phantom{00}
&3849.56\nl
8813.541 & 62.40\phantom{00} & 4077.23 && 9091.933& 79.35\phantom{00}
&3844.65\nl
8813.662& 92.40\phantom{00} &4075.28 && 9104.975& 92.40\phantom{00} &3938.08\nl
8813.681& 64.50\phantom{00} &4081.69 && 9104.992& 79.20\phantom{00} &3944.91\nl
8813.723& 92.40\phantom{00} &4080.03  && 9115.683& 93.30\phantom{00} &4042.42\nl
8813.742& 64.50\phantom{00} &4081.36 && 9115.698& 79.35\phantom{00} &4053.37\nl
8819.420& 92.25\phantom{00} &4064.21 && 9126.589& 91.35\phantom{00} &4098.67\nl
8819.438& 79.35\phantom{00} &4057.65 && 9126.603& 79.35\phantom{00} &4112.53\nl
8823.371& 92.25\phantom{00} &4075.47 && 9136.355& 91.35\phantom{00} &4120.11\nl
8823.390& 64.50\phantom{00} &4064.61 && 9136.369& 79.35\phantom{00} &4134.43\nl
8823.432& 92.40\phantom{00} &4030.04 && 9154.721& 82.35\phantom{00} &3963.77\nl
8823.451& 64.50\phantom{00} &4032.01  &&9154.736& 73.35\phantom{00} &3963.73 \nl
8827.522& 92.40\phantom{00} &4010.21 && 9163.753& 83.25\phantom{00} &3856.28\nl
8827.539& 79.35\phantom{00} &3993.68 && 9163.768& 73.20\phantom{00} &3909.88\nl
8830.469 & 92.25\phantom{00} & 3973.40  && 9210.749& 84.30\phantom{00}
&4136.98\nl
8830.479 & 64.50\phantom{00} & 3974.33 && 9210.767& 77.25\phantom{00}
&4137.28\nl
8830.529& 92.25\phantom{00} &3967.63  && 9227.202& 161.25\phantom{00}
&4133.59\nl
8830.548& 64.50\phantom{00} &4013.47 && 9227.214& 106.20\phantom{00} &4129.79\nl
8845.197& 92.40\phantom{00} &3993.66 && 9227.225& 106.20\phantom{00} &4160.07\nl
8845.216& 64.50\phantom{00} &4000.27  && 9237.104& 131.25\phantom{00}
&3989.11\nl
8845.259& 93.45\phantom{00} &4023.71 &&9237.116& 106.20\phantom{00} &3990.16\nl
8845.278& 64.50\phantom{00} &4013.82 && 9237.127& 104.40\phantom{00} &3989.45\nl
8845.330& 92.40\phantom{00} &4000.19 && 9265.502& 131.25\phantom{00} &3922.78\nl
8845.347& 79.50\phantom{00} &4003.18 &&   9265.514& 106.65\phantom{00}
&3965.31\nl
8855.975& 92.40\phantom{00} &4008.64 && 9265.525& 101.25\phantom{00}
&4032.79 \nl
8855.992& 79.35\phantom{00} &4004.59 && 9367.956& 131.25\phantom{00}
&4330.42   \nl
8857.844& 92.40\phantom{00} &4015.44 && 9367.968& 104.25\phantom{00} &4166.76 \nl
8857.862& 64.50\phantom{00} &4010.45 && 9367.979& 103.20\phantom{00}
&4102.59 \nl
%\tablebreak
9439.099& 160.50\phantom{00} &4029.86 && 10097.187& 712.05\phantom{00}
&4056.11\nl
9439.111& 104.25\phantom{00} &4032.52 && 10097.206& 657.75\phantom{00}
&4057.65\nl
9439.122& 104.25\phantom{00} &4035.69 &&  10118.897& 103.95\phantom{00}
&4146.12\nl
9448.744& 130.35\phantom{00} &4012.61 && 10118.907& 106.95\phantom{00}
&4153.15\nl
9448.756& 103.35\phantom{00} &4015.98&& 10118.911& 110.85\phantom{00}
&4154.69\nl
9448.767& 103.20\phantom{00} &4017.74 && 10147.225& 105.00\phantom{00}
&4054.12\nl
9458.461& 130.20\phantom{00} &4018.50 && 10147.235& 106.95\phantom{00}
&4058.27\nl
9458.473& 103.20\phantom{00} &4018.41 && 10147.239& 109.95\phantom{00}
&4097.83\nl
9458.484& 103.35\phantom{00} &4017.34 && 10208.929& 684.00\phantom{00}
&4186.41\nl
9467.510& 131.25\phantom{00} &4090.33 && 10208.944& 714.90\phantom{00}
&4178.43\nl
9467.522& 104.25\phantom{00} &4083.18 && 10208.959& 657.90\phantom{00}
&4174.19\nl
9467.529& 109.20\phantom{00} &4080.91 && 10263.709& 685.05\phantom{00}
&4038.82\nl
9476.629& 131.25\phantom{00} &4073.87 && 10263.726& 713.85\phantom{00}
&4039.39\nl
9476.641& 103.20\phantom{00} &4083.48 && 10263.745& 658.05\phantom{00}
&4038.92\nl
9476.652& 97.65\phantom{00}  &3651.69 &&  10335.487& 174.90\phantom{00}
&3929.83\nl
9924.038& 686.10\phantom{00} &3994.89 && 10335.498& 174.90\phantom{00}
&3929.85\nl
9924.055& 714.00\phantom{00} &4005.25 && 10335.509& 175.05\phantom{00}
&3935.00\nl
9924.074& 657.90\phantom{00} &4005.32  && 10335.514& 177.90\phantom{00}
&3936.73 \nl
9924.102& 107.85\phantom{00} &3990.95  && 10372.576& 172.95\phantom{00}
&4095.93\nl
9924.112& 106.95\phantom{00} &3995.77 && 10372.587& 173.85\phantom{00}
&4104.08\nl
9924.123& 106.95\phantom{00} &3992.63  && 10372.599& 170.85\phantom{00}
&4105.24\nl
9924.126& 109.95\phantom{00} &3996.53 &&  10372.603& 177.90\phantom{00}
&4112.94\nl
9951.202& 108.00\phantom{00} &4006.80 && 10556.684& 173.85\phantom{00}
&4077.49\nl
9951.212& 106.80\phantom{00} &4010.80 &&  10556.684& 173.85\phantom{00}
&4077.49\nl
9951.223& 108.90\phantom{00} &4032.56  && 10556.696& 170.70\phantom{00}
&4098.21 \nl
9951.226& 109.95\phantom{00} &4050.83  && 10556.696& 170.70\phantom{00}
&4098.21\nl
9985.447& 684.00\phantom{00} &4010.56 && 10556.707& 172.80\phantom{00}
&4083.36 \nl
9985.464& 712.20\phantom{00} &4003.36 && 10556.707& 172.80\phantom{00}
&4083.36\nl
9985.484& 657.90\phantom{00} &4023.77 &&   10556.711& 177.90\phantom{00}
&4089.84\nl
10000.199& 106.95\phantom{00} & 3971.31 && 10708.050& 172.90\phantom{00} &
4033.45 \nl
10000.209& 107.10\phantom{00} & 3968.96 && 10708.059& 176.90\phantom{00} &
4044.00\nl
10000.220& 102.90\phantom{00} & 3970.80 && 10708.070 & 174.70\phantom{00} &
4039.26\nl
10000.224& 109.95\phantom{00} &3974.83 && 10708.075 & 177.90\phantom{00} &
4039.48\nl
10082.549& 106.95\phantom{00} &3907.21 && 10736.169 & 172.90\phantom{00} &
3919.48\nl
10082.559& 106.05\phantom{00} &3910.19 && 10736.180 & 174.00\phantom{00} &
3936.21\nl
10082.570& 103.95\phantom{00} &3908.66 &&   10736.191 & 171.90\phantom{00}
& 3913.60\nl
10082.573& 110.10\phantom{00} &3916.68 && 10736.196 & 177.90\phantom{00} &
3912.12\nl
10097.170& 684.90\phantom{00} &4043.80
\enddata
\end{deluxetable}
\clearpage

\begin{deluxetable}{lrlllrl}
\tablewidth{0in}
\tablenum{A1.2}
\tablecaption{Barnard's Star Log of Observations} 
\tablehead{& \colhead{Exposure} & \colhead{Average} &
& &\colhead{Exposure} & \colhead{Average}
\\
\colhead{mJD} &
\colhead{Time} &
\colhead{Intensity} &&\colhead{mJD} & \colhead{Time} & \colhead{Intensity}}
\startdata
9025.634	&	80.4\phantom{000}	&	15065&&9439.92	&	39.45\phantom{00} 	&	15054.7\nl
9025.649	&	24.3\phantom{000}	&	15046.7 && 9439.927	&	35.25\phantom{00}
	&	15053.4\nl
9025.658	&	24.3\phantom{000} 	&	15064.8 && 9473.56	&	90.3\phantom{000}
	&	14989.1\nl
9088.103	&	50.25\phantom{00} 	&	15106.1 && 9473.575	&	38.4\phantom{000}
	&	15012.9\nl
9088.117	&	24.3\phantom{000} 	&	15116.5 && 9473.581	&	35.4\phantom{000}
	&	15022.7\nl
9088.127	&	24.45\phantom{00} 	&	15115.3 && 9487.853	&	37.35\phantom{00}
	&	15007.9\nl
9102.763	&	50.25\phantom{00} 	&	15069.3 && 9487.86	&	34.35\phantom{00}
	&	15012.2\nl
9102.777	&	24.3\phantom{000} 	&	15091.7 && 9506.676	&	91.2\phantom{000}
	&	14955.1\nl
9102.788	&	24.15\phantom{00} 	&	15093.1 && 9506.691	&	36.3\phantom{000}
	&	14945.1\nl
9118.631	&	51.15\phantom{00} 	&	14978.7 && 9506.698	&	33.3\phantom{000}
	&	14947\nl
9118.645	&	24.3\phantom{000} 	&	14994.9 && 9522.698	&	93.15\phantom{00}
	&	14870.4\nl
9118.658	&	24.3\phantom{000} 	&	14979.9 && 9522.713	&	33.15\phantom{00}
	&	14880.3\nl
9128.672	&	49.35\phantom{00} 	&	14895.8 && 9522.72	&	34.05\phantom{00}
	&	14886.2\nl
9128.686	&	24.45\phantom{00} 	&	14914.1 && 9536.508	&	92.1\phantom{000}
	&	14918.5\nl
9128.699	&	24.45\phantom{00} 	&	14915.4 && 9536.523	&	34.05\phantom{00}
	&	14901.1\nl
9131.751	&	49.35\phantom{00} 	&	14880.2 && 9536.53	&	33.9\phantom{000}
	&	14931.1\nl
9131.765	&	24.45\phantom{00} 	&	14895.7 && 9566.407	&	122.1\phantom{000}
	&	15001.8\nl
9131.777	&	24.3\phantom{000} 	&	14905.7 && 9566.422	&	34.05\phantom{00}
	&	14967\nl
9184.57	&	122.25\phantom{00} 	&	15045.9 && 9566.428	&	33\phantom{.0000}
	&	14977.1\nl
9184.585	&	33.3\phantom{000} 	&	15028.5 && 9581.286	&	122.1\phantom{000}
	&	15033\nl
9184.591	&	32.25\phantom{00} 	&	15034.9 && 9581.301	&	34.05\phantom{00}
	&	15016.3\nl
9216.568	&	91.35\phantom{00} 	&	15047.8 && 9581.306	&	34.05\phantom{00}
	&	15033.6\nl
9216.583	&	34.35\phantom{00} 	&	15021.6 && 9594.493	&	121.95\phantom{00}
	&	15007.1\nl
9216.586	&	33.3\phantom{000} 	&	15035.4 && 9594.508	&	37.05\phantom{00}
	&	15028.9\nl
9425.762	&	120.3\phantom{000} 	&	15082 && 9594.515	&	35.1\phantom{000}
	&	15050.2\nl
9425.777	&	35.25\phantom{00} 	&	15070.2 && 9609.505	&	121.8\phantom{000}
	&	14962.2\nl
9425.781	&	35.25\phantom{00} 	&	15080.4 && 9609.52	&	36\phantom{.0000}
	&	14965.5\nl
9439.905	&	91.2\phantom{000} 	&	15076.4 && 9609.527	&	35.85\phantom{00}
	&	14982.9\nl

9622.11	&	121.95\phantom{00} 	&	14952.7 && 9962.146	&	76.95\phantom{00}
	&	14996.1\nl
9622.126	&	36\phantom{.0000} 	&	14925.8 && 9962.157	&	75\phantom{.0000}
	&	15001.7\nl
9622.132	&	35.85\phantom{00} 	&	14957.6 && 9994.169	&	117\phantom{.0000}
	&	14977.5 \nl
9636.119	&	122.1\phantom{000} 	&	14943.7 &&9994.181	&	76.95\phantom{00}
	&	15020.3\nl
9636.134	&	37.05\phantom{00} 	&	14916.4 && 9994.192	&	75\phantom{.0000}
	&	15035.7\nl
9636.141	&	36\phantom{.0000} 	&	14928.4 && 10024.262	&	118.05\phantom{00}
	&	14974.8\nl
9653.682	&	121.95\phantom{00} 	&	14918.7 && 10024.274	&	78\phantom{.0000}
	&	14997.2\nl
9653.697	&	37.95\phantom{00} 	&	14923.1 && 10024.285	&	75\phantom{.0000}
	&	15002.1\nl
9653.703	&	34.95\phantom{00} 	&	14938 && 10117.955	&	13.2\phantom{000}
	&	14986\nl
9663.199	&	122.1\phantom{000} 	&	14946.8 && 10117.968	&	75.9\phantom{000}
	&	14986.7\nl
9663.214	&	37.95\phantom{00} 	&	14967.6 && 10117.979	&	48.48\phantom{00}
	&	14981.9\nl
9663.221	&	34.05\phantom{00} 	&	14984.8 &&10143.754	&	77.1\phantom{000}
	&	14982.8 \nl
9757.843	&	123\phantom{.0000} 	&	15002.6 && 10143.768	&	75.75\phantom{00}
	&	14978\nl
9757.858	&	36.15\phantom{00} 	&	14984.8 &&10143.778	&	75\phantom{.0000}
	&	14980.8 \nl
9757.864	&	33.9\phantom{000} 	&	14997.5 && 10167.077	&	115.95\phantom{00}
	&	14968.8\nl
9777.147	&	92.1\phantom{000} 	&	14937.5 &&10167.077	&	115.95\phantom{00}
	&	14968.8 \nl
9777.162	&	36\phantom{.0000} 	&	14910.5 &&10167.077	&	115.95\phantom{00}
	&	14968.8 \nl
9777.169	&	34.95\phantom{00} 	&	14929.3 && 10167.1	&	73.95\phantom{00}
	&	14985.5\nl
9916.557	&	116.85\phantom{00} 	&	14927.3 &&10198.508	&	116.85\phantom{00}
	&	14903.2\nl
9916.57	&	76.95\phantom{00} 	&	14925.7 && 10198.52	&	76.95\phantom{00}
	&	14921.4\nl
9916.581	&	75\phantom{.0000} 	&	14919.7 &&10198.531	&	75.9\phantom{000}
	&	14928.4 \nl
9935.324	&	117\phantom{.0000} 	&	14882.6 && 10237.311	&	118.05\phantom{00}
	&	14914\nl
9935.336	&	77.85\phantom{00} 	&	14917 && 10237.324	&	80.85\phantom{00}
	&	14942.5\nl
9935.347	&	76.05\phantom{00} 	&	14934.9 && 10237.335	&	56.73\phantom{00}
	&	14931\nl
9962.134	&	117\phantom{.0000} 	&	14984.8
\enddata
\end{deluxetable}